\documentclass[preprint2]{aastex}

\newcommand{\zsun}{{\,Z_\odot}}
\newcommand{\Teff}{T_{\rm eff}}

\accepted{1999, November 9}

\begin{document}

\title{Mid-UV Determination of 
Elliptical Galaxy Abundances and Ages}

\author{Jennifer M. Lotz\altaffilmark{1} }
\affil{Johns Hopkins University, 3400 N. Charles St., Baltimore, MD 21218 \\}

\author{Henry C. Ferguson\altaffilmark{2} and Ralph C. Bohlin\altaffilmark{3}} 
\affil{Space Telescope Science Institute, 3700 San Martin Drive, Baltimore, MD 21218 \\ accepted by the Astrophysical Journal}

\altaffiltext{1}{jlotz@pha.jhu.edu}
\altaffiltext{2}{ferguson@stsci.edu}
\altaffiltext{3}{bohlin@stsci.edu}

\slugcomment{accepted by the Astrophysical Journal}
\shorttitle{Mid-UV Determination of Elliptical Galaxy Abundances and Ages}
\shortauthors{Lotz et al. 1999}

\begin{abstract}
We investigate the effects of abundance and age on the mid-UV spectra and
Mg$_{2}$ strengths of stellar populations using simple population synthesis
models. These models are used to constrain the star formation history
of four nearby
elliptical galaxies and spiral bulges.  The mid-UV (1800 - 3200 \AA) light
of evolved stellar populations ($>$ 1 Gyr) is dominated by the main sequence
turn-off, unlike the optical light which is dominated by 
the red giant branch.  Because the main sequence turn-off is 
sensitive to age and metallicity in ways different from the RGB, 
a detailed investigation of the mid-UV
features of elliptical galaxies may help break the age-metallicity degeneracy
that plagues optical techniques. Also, a better understanding of this
wavelength region is useful for the studies of 0.5 $\leq$ z $\leq$ 1.5
 galaxies for which the
rest frame mid-UV is redshifted into the visible.
We create simple, single age (3-20 Gyr), 
single metallicity (Z $=$ 0.0004 - 0.05) spectral energy distributions (SEDs) 
extending into the UV using the Kurucz
model stellar fluxes.
Comparison to standard stars' mid-UV spectra reveals that the Kurucz
model fluxes accurately
model a blend feature of FeI and MgI at 2538 {\AA} (Bl2538)
and the slope of the continuum between 2600 and 3100 {\AA} ($S2850$).
We find that our simple single age, single metallicity SEDs agree well with
these mid-UV features of globular clusters.  However, the majority of the
 galaxies do not agree with the Bl2538, $S2850$, and Mg$_{2}$ values given
by these simple models.  
The mid-UV features appear to require 
both an old metal-rich and an old metal-poor
(Z $\leq$ 0.001) population.  The implied metal-poor population is less 
than 10\% of the total mass for all the galaxies but dominates the SEDs
shortward of 3000 {\AA}.  Intermediate age (1-5 Gyr) populations are not
required to match the UV for any of the galaxies, but are not ruled out. 
Despite being limited by the quality of the model stellar fluxes, our 
study has yielded two promising mid-UV spectral diagnostics (Bl2538 and 
$S2850$) and suggests unique and complex star formation histories for 
elliptical galaxies.
\end{abstract}


\section{Introduction}
The integrated light of elliptical galaxies contains many clues to
 their star formation histories.
Currently, the most popular approach to deciphering the broad band colors
and spectral features of elliptical galaxies and star clusters is 
evolutionary population synthesis (Charlot \& Bruzual 1991; Bressan, Chiosi, \& Fagotto 1994; Worthey 1994).  This technique models the
integrated light for a stellar population of a given
age and metallicity by summing up the light from individual stars along 
theoretical isochrones.  The resulting spectral energy distribution
(SED) is dependent on the libraries
of stellar evolutionary tracks and stellar fluxes, the chosen star
formation rate and initial mass function (IMF), and in the case of multi-metallicity populations,
the chemical enrichment history.  Charlot, Worthey, \& Bressan (1996)
examined the differences between several recent models and found that
they are seriously limited by the dearth of reliable libraries for cool
and non-solar composition stars and by the uncertain properties of
post main sequence stars.  Nevertheless, population synthesis is appealing
because of its ability to prescribe a fundamental recipe for a galaxy.

Attempts to unravel the history of star formation in elliptical galaxies via 
spectral synthesis are plagued by the ``age-metallicity degeneracy'' 
(Worthey 1994 and references therein).  The effective temperature of the red giant branch (RGB) decreases
with increasing age and metallicity in 
such a way that the RGB of an old solar metallicity population looks nearly
identical to that of a more metal-rich, slightly younger population. 
Because the RGB dominates both the integrated line strength and broad band 
color longward of 4000 {\AA}, 
a degeneracy of these properties in age and abundance results for stellar 
populations older than 1 Gyr. 
The
most promising attempts to disentangle age and metallicity from integrated
spectra have involved use of the Balmer absorption lines 
\citep{Gonzalez93, Greggio96} and mid-UV colors (Dorman, O'Connell, \& Rood 1995).
All of these methods rely on spectral diagnostics for which
the main sequence begins to dominate over the RGB.  

The contribution of the
main sequence turn-off to the integrated light is greatest in the mid-UV region ($\sim$ 2500-3500 {\AA}); the contribution of the far-UV bright HB and post HB
 stars to the integrated light is less than 50\% at 2500 {\AA}
(Worthey, Dorman, \& Jones 1996). 
\citet{HBHLYFGLMSKLTLBBCGLK98} suggest that the 
spectral break at 2900 {\AA} is correlated with spectral type of the turn-off
population but
weakly influenced by metallicity. 
Also, the continuum of mid-UV is affected by the metallicity of the 
stellar population.  \citet{DOR95} found that the color (2500-V) was 
correlated with metallicity and Fanelli et al. (1990, 1992) 
 found $\delta$ (2600-V)
 10 times
more sensitive to [Fe/H] than $\delta$ (B-V). Because the spectral type of 
the main
sequence turn-off stars dominates the absorption features 
of the mid-UV and the 
continuum is highly sensitive to chemical content, a detailed investigation
of the integrated mid-UV light of stellar populations may help break the
 age-metallicity degeneracy.  

One observational result population synthesis models have sought to explain is the correlation
between the ``UV upturn'' shortward of 2000 {\AA} and the optical Mg$_{2}$ absorption line strength in
elliptical galaxies \citep{Worthey94, DOR95, BCF94}.  The brightest galaxies in UV 
also have the strongest Mg$_{2}$ absorption (Burstein et al. 1988), 
suggesting that these UV strong galaxies are
metal-rich and  possibly enhanced in $\alpha$ element abundances.  
In the most metal-rich galaxies, the primary source of far-UV emission appears
to be extreme horizontal branch (EHB) stars and their descendants, the 
AGB-manqu\'e\ stars.  While the bulk of the evidence favors a metal-rich
origin for the EHB stars \citep{DOR95, BFDD97}, models
with a distribution of metallicities including a metal-poor tail
 \citep{PL97} cannot be 
excluded.  In any case, the models for hot populations are strongly dependent
on mass loss on the red giant branch and on the helium abundance.  Small
changes in these two parameters can give wildly different UV upturns at fixed
age and metallicity.  For the purposes of this paper, the UV upturn is simply
an underlying continuum that must be subtracted from the galaxy spectra prior
to analysis of the spectral features.

The star formation history of nearby ellipticals,
particularly M32, has been in the subject of much debate.
In 1980, \citet{O'Connell80}  proposed that a 
5 Gyr stellar sub-population is responsible for the relatively blue
$B-V$ color of M32.  The idea that M32 contains a large intermediate
age population was supported by detailed comparison of M32 and 47 Tuc (an
old ``metal-rich'' globular cluster)
 near UV spectra (3800 - 4400 {\AA} ) \citep{Rose94}. The study of
H$\beta$ and H$\gamma$ absorption lines, which are dependent on the main 
sequence turn-off stars in old stellar populations, suggest intermediate
ages for many ellipticals \citep{Gonzalez93, Greggio96}. However, 
recent HST observations of M32's red giant 
branch implies that at least half of M32's stars are older than 8 Gyr  
\citep{GLWFFMABHLOS96}.  The metallicity distribution of M32 and
other ellipticals has also been investigated. \citet{RD99} 
found evidence for a hot  metal-poor population in an
otherwise metal-rich population in the mid-UV spectra of M32. 
The width of M32's RGB requires that the stars
span a substantial range in metallicity 
\citep{GLWFFMABHLOS96}.  The estimated metallicity distribution has
a low metallicity tail extending down to [Fe/H] = -1.5 but is
strongly skewed to  solar metallicity and is much narrower than
expected from a closed box chemical enrichment scenario.  Comparisons to
synthetic spectra suggest that other
galaxies, including NGC 1404, NGC 4649, and the bulge of M31, also show
a deficit of low metallicity stars relative to closed box models 
\citep{BCF94, WDJ96}.

The ages of distant elliptical galaxies has been the subject of some debate
as well.  \citet{SDSDPJW97} determined an age of $\sim$ 3.5 Gyr (assuming a
solar metallicity) for LBDS 53W091 (z $\sim$ 1.55) by comparing the rest
frame mid-UV spectral features to solar metallicity F and G stars and 
population synthesis models. A galaxy this old at a redshift of 1.55 would
rule out a $H_0 > 50 $, $\Lambda = 0 $, $\Omega = 1$ universe.
A later comparison of this galaxy's spectra to improved F star models 
gave an age of 1-2 Gyr (assuming Z $\geq$ $\zsun$), relaxing the cosmological
constraints (Heap et al. 1998).  Because galaxies with 0.5 $\leq$ z $\leq$ 1.5
 have their mid-UV spectra redshifted into the visible range, a better
understanding of the mid-UV features of old stellar populations 
will be useful for dating distant galaxies.

Using simple population synthesis models, we examine the mid-UV 
features and Mg$_{2}$ strengths of four nearby elliptical galaxies
 and spiral bulges. 
In order to study the spectral properties of the underlying main
sequence and RGB branch stars, we exclude the uncertain 
hot HB and post HB stars from our models' isochrones and attempt to 
subtract out the far-UV component from the galaxies' spectra.  Specifically
we use the blend feature at 2538 {\AA} (Bl2538) and Mg$_{2}$ line indices and introduce a new tool for 
examining the mid-UV continua, the slope between 2600 and 3100 {\AA} ($S2850$),
to determine the galactic properties.
The investigation of the mid-UV dependence on age and abundance via population
synthesis is severely limited by inaccuracies in our model stellar fluxes.
However, Bl2538 and $S2850$ appear to be adequately
modeled by the Kurucz fluxes.
We find for our simple population models that populations older than 3 Gyr 
separate out in the
S2850 v. Bl2538 and S2850 v. Mg$_{2}$ planes. Although
globular cluster spectra agree with simple, single metallicity, single age
populations, a small but mid-UV bright component of low metallicity (Z $\leq$
0.001) 
stars may explain the spectral features of NGC 1399.
M32's UV features cannot be explained purely by a significant 
population of intermediate age (1-3 Gyr) solar metallicity stars; a small
number of metal-poor stars are suggested by the mid-UV in this case as well.
Because metal-poor stars are so much brighter than metal-rich stars in the
mid-UV, we cannot ignore the effects of the metallicity distribution on our
population synthesis models. 

The rest of the paper is arranged as follows: first, in section 2 
we test our library
of stellar fluxes (Lejeune, Cuisnier, \& Buser 1997) for inaccuracies and determine the
influence of these inaccuracies on our model spectral energy distributions
(SEDs).  In section 3, we explain our population synthesis models, 
their mid-UV and optical properties, and compare them to the mid-UV features
of several composite globular cluster IUE spectra (Bonatto, Bica, \& Alloin 1995). 
In section 4, we present the FOS spectra of NGC 1399 and NGC 3610 and the 
IUE spectra of NGC 1399, M32 and the bulge of M31.  We compare our models to
these galaxies.  In section 5 we examine the
effects of a metal-poor sub-population on the integrated spectra of 
an otherwise metal-rich population.

\section{Testing Kurucz Stellar Fluxes}
  
In order to examine the effect of abundance on the integrated light of old
stellar populations, Kurucz model stellar fluxes \citep{LCB97}
 with [Fe/H] = -1.5, -1.0, -0.5, 0.0, and +0.5 
were used for our population synthesis models. 
Due to the paucity of non-solar abundance stars in empirical libraries, the 
Kurucz models offer the advantage of the studying the effects of a
full range of metallicity on the model stellar populations.
However, at cooler temperatures the model fluxes do 
suffer from serious limitations.  \citet{MFMKB93}
 compared late G and K 
solar chemical composition giants and found that, while the optical and 
infrared observations matched the Kurucz fluxes, significant discrepancies
existed in the UV.  They concluded that the models failed to account for
chromospheric emission and thus underestimated the flux at wavelengths
less than 3100 {\AA}.  Also, the optical Mg$_{2}$ index of cool,
 metal-rich giants may not be well modeled by the Kurucz stellar fluxes.
When the Kurucz 1992 models were compared to the improved 
Kurucz 1993 models, Chavez, Malgnini, \& Morossi (1995) found that for $\Teff$ $\leq$ 5000K and
[Fe/H] $\geq$ 0.0, the newer models predicted a Mg$_{2}$ values up to 
0.58 magnitudes stronger than the 1992 models.  The Kurucz 1993 fluxes 
corrected an error in the FeI continuum opacity for cool stars, had more
optical depth points, and included
detailed modeling of scattering effects in the source function 
 not accounted for in 
the 1992 fluxes. 
Clearly, while the Kurucz models are powerful
tools for the modeling of stellar populations, a detailed
 understanding of their
limitations is necessary to interpret the 
results of our model stellar populations correctly.

Here we quantify the accuracy of the Kurucz model stellar 
fluxes in the mid-UV and
the optical. We test the
mid-UV spectra and optical features Mg$_{2}$ and H$\beta$ of the solar 
abundance Kurucz 1992 models against the composite
IUE stellar main sequence spectral energy distributions complied by 
\citet{FOBW90} and standard stars in the 
\citet{GS83} catalog. The importance of the cool 
giants on the mid-UV and optical portions of our model SEDs is determined.

\subsection{$U-B$ v. $B-V$ for dwarfs and giants}

We have used the un-calibrated Kurucz model fluxes presented by
\citet{LCB97} as our stellar library.  Lejeune et al. (1997) have calculated
correction functions for each of the Kurucz models in order to 
yield synthetic broad-band colors which match the empirical color-temperature
relations.  However, they note that the
dwarf stellar fluxes have a negligible optical 
color correction for $\Teff$ $\geq$
3500 K, and that the color correction is significant for the giants
only for $\Teff$ $\leq$ 4000 K.  Because we use the non-color corrected
fluxes, we first tested the un-calibrated solar metallicity model colors
against the colors of the stars in the Gunn-Stryker catalog \citep{GS83}.
The broad band colors U-B and B-V of the Kurucz fluxes  
match both the giant and dwarf stars in the Gunn-Stryker catalog to within
$\pm$ 0.2 magnitudes  (Figure 1a \& b). The coolest red giant colors are least
well fit by the Kurucz fluxes, with a discrepancy of $\sim$ 0.2 magnitudes.

\subsection{Mid-UV Spectral Indices}

Fanelli et al. (1990, 1992) 
examined the mid-UV spectral 
morphology of 218 normal stars observed by IUE and found that the 
 spectral lines in the UV
are sensitive to and change significantly
with spectral types ranging from O to K.  
We measured the spectral indices
Fe2402, Bl2538, Fe2609, Mg2800, Mg2852, MgWide, and Fe3000 
as defined by \citet{FOBW90} (Table 1 and Figure 2)
for the Kurucz model fluxes corresponding to
solar, main-sequence (MS) stars and compared them to the indices of the IUE
stellar spectra measured by \citet{FOBW90}. Unfortunately, few non-solar
metallicity stars were observed, and so our discussion is 
limited to the comparison
of solar abundance stars and the corresponding Kurucz models.  It is important
to note that the ``continuum'' band passes for these indices contain very little
true continuum.  Due to the strong line blanketing in the UV, we are forced 
to use pseudo-continuum band passes around the absorption features to compute
indices and our results will depend on the accuracy of the pseudo-continuum
as well as that of the absorption features of the stellar models.

In Figure 3a-g, we plot the
B-V color against the mid-UV line indices for both the IUE composite spectra
(triangles) and Kurucz fluxes (connected crosses).
The B-V colors for the Kurucz models were
measured using IRAF task synphot.calcphot. 
Only the Bl2538 feature is well matched (to better than 0.05 magnitudes) 
by the full temperature range
of Kurucz main sequence model fluxes (Figure 3b). (When
the individual IUE spectra are compared to the Kurucz fluxes, the
stellar Bl2538 values show a great deal of scatter about the Kurucz values
(Dorman, O'Connell, \& Rood 1997).  This may be due to noise in the
individual IUE spectra or metallicity and gravity variations. 
The difference in Bl2538 between a G0V and G0III star is $\sim$ 0.4 magnitudes;
 the difference between a G0V and G0IV star is much less, $\sim$ 0.03 
magnitudes. We examine
the composite IUE spectra of stars identified as
luminosity class V because these stars will dominate the mid-UV light.)
In general, the Kurucz model fluxes for B-V $\leq$ 0.1 ($\Teff$ 
$\geq$ 8000 K) match the observed
line strengths of IUE data, but overestimate the line strengths for
cooler stars (B-V $\geq$ 0.2, $\Teff$ $\leq$ 7000 K).  In particular, 
the $\Teff$ = 6000 K, log g=4.5 solar
Kurucz model is too strong by 0.6 magnitudes for Fe2404, 0.4 for Fe2609, 
0.4  for Mg2800,
0.4  for Mg2852, 0.3 for MgWide, and 0.15 too weak for Fe3000. 
These discrepancies may be due in part to the inevitably 
incomplete atomic line lists. They may also be the result of 
departures from LTE, chromospheric emission 
filling in the mid-UV absorption
lines for cooler dwarfs, and opacity due to tri-atomic molecules, none of which
are modeled by the Kurucz stellar fluxes.

\subsection{Mid-UV Continuum}

The continuum of the mid-UV range is strongly sensitive to 
the metallicity of the stellar population \citep{FOBW90, DOR95} 
Age may also influence the mid-UV continuum: the spectrum of an old stellar population at 2500 {\AA} is dominated by
 flux from 
main sequence turn-off stars, and thus may be used to constrain this 
turn-off population more closely than the K and M giant dominated 
optical region of the spectrum.  Again, it is important to realize that
the vast number of atomic absorption lines in the UV suppress most of the true
continuum and what we observe is a pseudo-continuum.

To examine the dependence of this region's pseudo-continuum 
 on age and metallicity, we
have measured the slope of the pseudo-continuum ($\Delta$ magnitude/1000 {\AA} ) 
in the 2600 to 3100 {\AA} range of our model simple populations (section 3).
The slope of the pseudo-continuum between 2600
and 3100 (hereafter $S2850$) was estimated using the IRAF V2.11.1 task 
onedspec.continuum with high 
rejection set at 3 $\sigma$ and low rejection set at 1 $\sigma$ in order to
exclude the effects of the magnesium features at 2800 and 2852 {\AA} on 
the spectral shape.  
Again, to determine possible inaccuracies in the model fluxes, we tested
$S2850$  of the Kurucz models against
$S2850$  of
the IUE observations.  In Figure 3h, the $S2850$ for the main sequence IUE
spectra and Kurucz models are plotted against $B-V$ for the range of slopes
observed in the galaxies.  
The Kurucz models' pseudo-continua
closely match that of the observations (to better than 0.5 $\Delta$ magnitude/
1000 {\AA} ), except for the coolest stars ( $(B-V)
\leq$ 1.1).  (Note: the color-corrected Lejeune et al. 1997 version of the
stellar models do not reproduce S2850; the shifts in B-V are small but
the S2850 feature is shallower by up to 2 mag/ 1000 {\AA}.)

\subsection{Mg$_{2}$ and H$\beta$}

We also examine the robustness of the Kurucz models' line
strengths at
optical wavelengths for solar metallicity giants.  
The Mg$_{2}$ feature is primarily a measurement of the 
galaxy's metallicity, but is also strongly linked to age.
The H$\beta$ feature at 4860 {\AA}  is sensitive to main sequence turn-off
temperature, and hence may be an indicator of age \citep{BCT96, Gonzalez93}. Observed elliptical galaxy H$\beta$ and H$\gamma$ absorption strengths
suggest intermediate ages ($\leq$ 8 Gyr).  
We seek to determine the relationship between the mid-UV 
properties and these optical line strengths for elliptical galaxies (section
4). 

In Figure 4a-d, we compared
 the Mg$_{2}$ and H$\beta$ features of the Kurucz solar MS dwarfs and giants 
models to a 
number of standard MS dwarfs and 
giants taken from the Gunn-Stryker atlas. 
The Kurucz models have an H$\beta$ equivalent width $\sim$ 1-2 {\AA} less than
the standard stars, so we do not use this optical feature for our analysis. 
Although  Mg$_{2}$ is well modeled by the Kurucz models with $\Teff$ $\geq$
4500 K, cooler giants increasingly overestimate the feature 
by 0.2 magnitudes or more
 (Figure 4a).
The coolest dwarf models also overestimate Mg$_{2}$ strength.
Given that \citet{CMM95} found that the cool metal-rich giant Kurucz
models' Mg$_{2}$ values were strongly sensitive to small changes in
the scattering parameter, we will also not use the Mg$_{2}$ values
estimated by the Kurucz models. Instead we use the empirical 
fitting formula given by \citet{BIdFPT95} (the dashed line
in Figure 4a,b). \citet{BIdFPT95}
determined the dependence of $\sim$ eighty stars' Mg$_{2}$ on effective 
temperature, gravity, [Fe/H], and [Mg/H].

\subsection{Importance of cool giants on model SEDs}
Our model SEDs were created using the technique of ``isochrone synthesis'',
whereby a stellar population is created by summing up stellar fluxes
along isochrones for a given age and metallicity; in section 3 we describe 
this process in greater detail.  Comparing the isochrones of different ages
and metallicity \citep{BBCFN94}, we see that old, metal-rich populations possess a significant
number of K and M giants.  Since the Kurucz stellar fluxes poorly model these
stars in the UV, we examined the importance
of the cool giants to the total integrated light of our oldest, most
metal-rich (20 Gyr, Z=0.05) model population. 

We have removed the contribution from giant stars with $\Teff$ $\leq$ 4500 K
from a 20 Gyr, Z=0.05 SED and compared it to the un-modified SED (Figure 5).
We see that the giants contribute very little light
to the mid-UV (1800-3200 \AA) , and become significant for wavelengths $\geq$
4000 {\AA}.  Thus we are confident that the mid-UV continuum of our synthetic populations
is not significantly affected by the known inaccuracies 
of cool giants.  

In summary, after close investigation of the mid-UV  and optical line
strengths of the main sequence and giant
 solar metallicity Kurucz models, we are left with
only the blend feature of FeI and MgI 
 at 2538 {\AA} in the mid-UV ($Bl2538$) and  the slope of the 
continua between 2600 and 3100 {\AA}  ($S2850$) 
as features reproduced well by the Kurucz models.
We are satisfied that the mid-UV is not significantly
influenced by the discrepancies in the cool giant 
stellar models.  
However, if further detailed investigation of the narrow 
band absorption features in the mid-UV is to be done, 
new model stellar fluxes with 
more accurate mid-UV and optical 
line strengths need to be constructed.  It is important to note that our
tests of the Bl2538 and $S2850$ features have only shown that the relative
fluxes of the pseudo-continuum over small wavelength regions are reasonably
matched by the Kurucz models.  For Bl2538, we have only examined the relative
flux of the local pseudo-continuum to the blend absorbtion feature, and 
for $S2850$, we have tested the average slope of the pseudo-continuum over a
500 \AA\ range.  A 10 \% error in the mid-UV pseudo-continuum would introduce
a 0.1 magnitude error in  Bl2538 and a 0.2 magnitude/ 1000 \AA\ error in 
$S2850$.
Problems with the mid-UV pseudo-continuum over
larger wavelength regions and its normalization relative to the optical 
pseudo-continuum may still exist.

\section{Single metallicity stellar population models}
\subsection{Spectral synthesis}
Single metallicity, single age simple stellar populations were constructed
using the ``isochrone synthesis'' technique, which sums up model stellar
fluxes using a given set of isochrones, IMF and star formation rate.
We used Kurucz model fluxes for stars with 3500 K $\leq$ $\Teff$ $\leq$
40000 K, the Padova isochrones \citep{BBCFN94}, a Salpeter IMF, 
and an exponentially decaying star 
formation rate with $\tau$ = $10^{7}$.  Spectral energy distributions were created for
Z=0.0004, 0.001, 0.008, 0.02, and 0.05 and ages 3, 7, 10, 12, 15, and 20 Gyrs.
Further details of the isochrone synthesis code are given in \citet{BF96}.

The Padova isochrones only contain ``normal'' P-AGB stars and do not 
model post-early AGB (PE-AGB), AGB-manqu\'e\ or ``extreme blue horizontal 
branch'' stars (HB stars with ZAHB $\Teff$ $\geq$ 16,000 K) 
which are all possible contributors to the
UV upturn found in elliptical galaxies \citep{BFDD97}.  
While the existence of EHB stars and their descendents in elliptical galaxies
is reasonably secure observationally, theoretical models of 
how the EHB is populated
are uncertain.  The number of EHB stars depends on parameters such as the
metallicity, age, helium abundance of the stellar population, and on 
the details of mass loss at the tip of the red giant branch.  
While it is possible to choose plausible values for these parameters and 
construct synthetic spectra, the dependence on the poorly constrained RGB
mass loss rate means that the UV upturn is not yet useful for 
constraining the age and chemical content of the stellar population.

For galaxies with a strong UV upturn, the mid-UV portion of the spectrum 
contains flux both from EHB stars ( $\Teff$ $\sim$ 25,000 K) and from
cooler HB and MS turn-off stars. Because we are interested in the MS turn-off
stars, we have chosen to subtract the hot  stellar 
component from both the models and the data (section 4).
The isochrones were modified
to eliminate HB and P-AGB stars with $\Teff$ $\geq$ 9,000 K (Figure 6).
With these modified tracks, spectral energy distributions were created for
Z=0.0004, 0.001, 0.008, 0.02, and 0.05 and ages 3, 7, 10, 15, and 20 Gyrs.

\subsection{Mid-UV and Mg$_{2}$}
In section 2, we evaluated the accuracy of the Kurucz models and found that
the blend feature at 2538 {\AA}  and the slope $S2850$  were well 
modeled.  We measured Bl2538 and $S2850$ for the single age (3 - 20 Gyr), 
single metallicity (Z=0.0004 - 0.05) 
models.  Mg$_{2}$ was calculated using the empirical formula from
\citet{BIdFPT95}:
\begin{eqnarray}
Mg_{2} = \exp\{-9.037 + 5.795 \cdot \theta + 0.398 \cdot \log g \nonumber \\
 + 0.389\cdot [Fe/H] - 0.16\cdot [Fe/H]^{2}  \nonumber \\
+ 0.981\cdot [Mg/Fe]\},
\end{eqnarray}
where $\theta$ = 5040/$\Teff$ (plotted as a dashed line in Figure 4a,b).
Because the Kurucz model stellar fluxes do not model 
the effect of enhanced $\alpha$ element abundance
on the mid-UV spectrum, we assumed [Mg/Fe] = 0.  An enhanced [Mg/Fe] has
a strong effect on Mg$_{2}$ at the higher metallicities; a 15 Gyr,
 [Fe/H]=0.0, [Mg/Fe]= +0.5 population has a Mg$_{2}$ feature 0.15 magnitudes
stronger than a similar [Mg/Fe]=0.0 population. 

In Figures 7-9 we plot the relations between $Bl2538$, $S2850$, and Mg$_2$
as a function of age and metallicity.  The models generally occupy a narrow
locus on these figures, indicating an age-metallicity degeneracy for simple
stellar populations.  However, young metal-rich populations do not
overlap with old metal-poor populations due to 
the curvature in the relations shown in Figures 7 and 9.  This curvature
also means that populations with a spread in metallicity will not follow
 the simple stellar population locus.  We will return to this in section 5.

\subsection{Testing Models against Globular Clusters}

The tests of the Kurucz model fluxes in section 2 were valid only for solar
metallicity.  In this section we compare the features predicted by the simple
stellar population models to the observed features in globular clusters
to test the synthesis model's validity at low metallicity and old ages.
\citet{BBA95} published mid-UV equivalent 
widths (EW) and continuum
point relative flux values normalized to the flux at 2646 {\AA} for 
composite globular cluster IUE spectra (Table 2).  We chose the
seven composite spectra for globular clusters with ages $\geq$ 10 Gyr and
metallicities Z= 0.0002 to 0.006 to test the validity of the slopes and 
mid-UV Bl2538 features of our simple models. We are not able to estimate
the hot stellar component of the integrated globular cluster spectra because
the full spectra are not given.  Therefore we computed our model SEDs with 
the original isochrones (including the hot HB and post HB stars) and compared
these to the composite globular cluster features (Figure 10).  
The presence of the hot
stellar component has little effect on the low metallicity models, but 
significantly changes the Bl2538 feature for the Z $\geq$ 0.02 models.

\citet{BBA95} measure the equivalent width between 2506-2574.  This is
approximately 40 {\AA}  wider than the Bl2538 index, and encompasses
SiI absorption lines at 2507, 2514, 2516, thus translating this EW into
an index may cause a slight overestimation of the
strength of the Bl2538 index.  
Fitting the slope of the globular cluster spectra was relatively straight-forward.  The slope of a line fit to the log of the flux at the continuum
points 2646, 2959, and 3122 {\AA} was measured.  In order to convert
 the equivalent widths
to indices, the flux of the continuum for the EW 
bandpass 
at 2556 {\AA} ($C_{2556}$) was interpolated between the continuum flux at
2466 {\AA} and 2646 {\AA}. Then the ``area'' of the absorption 
feature was determined by 
multiplying the EW by this interpolated continuum flux $C_{2556}$.
Then the depth
of the absorption feature was estimated by dividing this area by the width
of the EW bandpass (68 {\AA} ).  Finally the flux of the absorption feature 
at 2556 {\AA}  ($F_{2556}$) was calculated by 
subtracting the depth of the absorption feature from $C_{2556}$ and the 
Bl2538 index value was estimated by -2.5 log ($F_{2556}$/$C_{2556}$).

Table 2 gives the EW and continuum values measured by
\citet{BBA95}, and the calculated indices and slopes.  The calculated indices 
and slopes have large uncertainties, since these values are only rough
estimates from the data.  Nevertheless, plotting the globular cluster points
(triangles)
on Figure 10, we find good agreement with our simple SEDs, except for
47 Tuc.  The
difference between the globular clusters (except for 47 Tuc) and the models
is $\leq$ 0.05 in $Bl2538$ and $\leq$ 0.35 in $S2850$. When the EW of 
our Z=0.0004, 0.001, and 0.008, 15 Gyr model SEDs were measured, 
they agree well with the globular cluster EW although the model Z=0.008 EW
is slightly less than the clusters' EWs.  
The globular clusters'
slopes and Bl2538 strengths increase with metallicity, as predicted by our
single age, single Z models. 
 
Note that the cluster slopes are slightly lower than the model 
predictions (although within the uncertainties), which may be due 
an underestimation of the slope of the clusters because of uncertainties
in the hot HB and post HB stellar
component. The Padova isochrones do not possess extreme horizontal
branches, which may affect the mid-UV light of some of the globular clusters.  
Note the difference between the the ``blue'' Z=0.0007 cluster
and the ``red'' Z=0.0006 cluster in Bl2538.  Again, this may be the result of
the blue component of stars washing out the Bl2538 feature.

\subsection{Reddening effects}
The effects of interstellar extinction on the $S2850$ and Bl2538 were
found to be less than the uncertainty of the model SEDs and galaxy data.  
Galactic extinction curves with A$_{B}$ = 0.0, 0.05, 0.10, 0.25, 0.50, and
0.75 \citep{Pei92} were used to redden the SEDs of solar metallicity models
of ages 1, 3, 7, 10, 15, and 20 Gyr and the SEDs of a 7 Gyr population
with Z= 0.0004, 0.001, 0.008, 0.02, and 0.05. 
 In all cases, $\Delta$ Bl2538 $\leq$
0.05 magnitudes.  For A$_{B}$ $\leq$ 0.50,  $\Delta$ $S2850$ $\leq$ 0.4 and
was 0.6 for A$_{B}$ = 0.75. 
Thus, for Galactic reddening with A$_{B}$ $\leq$ 0.50, the reddening effects
on the slope and the blend feature are less than the error bars on the 
galaxy data points (Figure 7).

\section{Galaxy spectra}

\subsection{FOS observations of NGC 1399 and NGC 3610}

NGC 1399 and NGC 3610 were observed with the HST Faint Object Spectrograph
on 1993 17 January and 1992 30 November, respectively (Figure 11). Gratings and exposure
times are shown in Table 3.
All observations used a 1.0 \arcsec \ diameter circular aperture. The observations
were with the aberrated HST optics, prior to installation of COSTAR. 
In this paper we focus on the G270H data. 

The target acquisition procedure involved measuring the flux a 4x4 
grid for spaced at 0.25" intervals, and choosing the brightest spot
in the grid. This procedure thus centered the galaxy to an accuracy
of 0.25" in the aperture, and also resulted in the observations with
the red channel and the blue channel being slightly offset from 
each other (in retrospect a 5x5 grid would have been better).

The FOS data were reduced following standard pipeline procedures. 
This involves correcting for geomagnetic-induced image motion;
correcting for particle-induced dark current 
based on the geomagnetic position of the spacecraft; correcting 
for scattered light via subtraction of a constant 
count-rate determined from diodes that are unilluminated by the selected 
grating; flat-field correction; wavelength calibration; and flux
calibration based on observations spectrophotometric standards. 

A full correction for grating-scattered light is in 
principle much more complex than the simple subtraction of
a constant we have adopted here (Bushouse, Rosa, \& Mueller 1995).
 In particular, due to the high sensitivity of
the FOS detectors to long-wavelength photons, the FOS spectra contain
a component of light scattered from wavelengths longward of 3000 {\AA}.
We have used bspec \citep{Rosa94} to estimate the uncertainty
introduced by ignoring the wavelength dependence of the scattered
light correction. For NGC1399 we use a composite spectrum constructed
from the the HUT spectrum, the IUE spectrum of NGC4649, and an 
optical spectrum from the Las Campanas Observatory 
(Kimble, Davidsen, \& Sandage 1989), 
normalized to agree in the wavelength regions where they overlap.
For this model, the average scattered light contribution to the measured flux
from 1700-3270 {\AA} is less than 1\%.

\subsection{IUE observations}
We have obtained  UV spectra for NGC 1399, M32 and the bulge of M31,
 extracted from the IUE database by D. Calzetti (Figure 11).  
All the spectra have been corrected for redshift obtained from NED 
using the IRAF V 2.11.1 task onedspec.dopcor.  
Literature values for Mg$_{2}$ 
(Burstein et al. 1988; Worthey, Faber, \& Gonzalez 1992) may 
overestimate the galaxy's mean Mg$_{2}$ index strength because the 
spectra are often
taken with a narrow aperture centered on the
inner-most (and probably most metal-rich) regions of the galaxy.  We
measure the Mg$_{2}$ values from the optical spectra of \citet{KCBMSS96} which were taken with an aperture matched to the 10\arcsec \ $\times$ 
20\arcsec \ IUE aperture.  For NGC 1399, 20\arcsec \ corresponds to $\sim$ 2
kpc.  However, for M 31 and M 32, 20\arcsec \ corresponds to $\sim$ 80 pc, so
for these galaxies even with the wide aperture slit we are observing 
only the very central regions.

In the case of NGC 3610, wide aperture optical spectra were not available
and the Mg$_{2}$ value taken with the Lick 1.4\arcsec $\times$ 4\arcsec
\ aperture \citep{WFG92} was compared to our models. If we assume NGC 3610
has a metallicity gradient typical of other elliptical galaxies \citep{DSP93, CDB93}, no appreciable age gradient, and an age
$\sim$ 10 Gyr, at a 10\arcsec \ radius, NGC 3610 should have a slope 0.5 
magnitudes/1000 \AA \ less steep, Bl2538 0.05 magnitudes weaker, and an Mg$_{2}$
0.05 magnitudes weaker than measured in a 1\arcsec \ aperture.

\subsection{Hot component subtraction}
Our goal is to examine the main sequence and cooler stellar sub-population of
these galaxies, and not the hot horizontal and post HB populations. We must
therefore determine which galaxies possess significant UV upturns and 
attempt to 
subtract away this hot component of the spectra. We estimated the strength 
of the UV upturn by measuring the $1550-2500$ color of the 
galaxies with the IRAF
task onedspec.sband.  The 1550 and  2500 bands were centered at 1550 
and 2500 {\AA}  respectively and were both 600 {\AA}  wide.  
Our calculations of $1550-2500$ agreed with previously published values for the
galaxies. 

M32  had the reddest $1550-2500$ color ($\sim$ 1.5).
The rest of the galaxies have significant UV upturns indicative of a hot stellar
population. \citet{FD93} found that the
far-UV flux of M31 shows a drop-off blue-ward at 1000 {\AA}, which requires that
the $\Teff$ be $\sim$ 25,000 K for the UV emitting stars. 
Thus, for the ``normal'' galaxies with no recent star formation, the
UV upturn can be roughly approximated by a $\Teff$ $\sim$ 20000 - 30000 K
star. Between 1350 and 3200 {\AA}, the spectral shape changes very little
with temperature for stars hotter than 20,000 K.
 For the galaxies with 1550-2500 colors $\leq$ 0.5 magnitudes, 
we fit a 20,000K log g=5 Kurucz model to
the galaxies' flux between 1350 and 1650 {\AA}  and subtracted the stellar
spectrum from the galaxies spectra (Figure 12).  To test whether this method did a good
job of eliminating the light from the hot component, models of a 
simple single metallicity population where created using the original 
isochrones with hot HB and P-AGB stars.  Then the 20,000 K stellar model 
was fit
and subtracted from this model SED.  The resulting spectra had  
an identical slope and Bl2538 strength as the model created using the modified
isochrones (section 3.1).  
Thus if a galaxy's UV upturn is due to a $\sim$ 20,000 K hot stellar component,
the subtracted galaxy spectra should have $S2850$ and Bl2538
which reflect the metallicity and ages of the cooler stars.

If a significant population of ``blue
stragglers'' with $\Teff \sim 10,000K $ exists, the resulting spectra will 
still have contamination from this non-MSTO population.  However, like the
hot HB and post HB population, the blue straggler population in galaxies
is not well constrained.  Spinrad
et al. (1997) estimate that if elliptical galaxies have blue straggler
 populations similar to that of old metal-poor clusters, blue stragglers may 
contribute as much as 20-50 \% of flux at 2600 {\AA}.  On the other hand, 
for the 
higher metallicity cluster 47 Tuc, the blue straggler contribution to the
mid-UV flux has been estimated to be 7\% \citep{RD99}.

\subsection{General trends in spectral features}
Figures 7-9 show the measured spectral indices for the four galaxies in our
sample.  While the sample is small and the error bars are large, there 
are at least hints of some general trends.  In particular, as Mg$_2$ increases,
the $Bl2538$ feature tends to get weaker and the slope $S2850$ tends to 
get shallower.  This is clearly evident in the comparison of NGC 1399 (Mg$_2$
= 0.312) and NGC 3610 (Mg$_2$ = 0.271) in Figure 11.  It is also clear
in Figure 7-9 that the galaxies do not follow the behavior expected for simple
stellar populations.  The galaxies are generally displaced from the
model locus and the trends are nearly orthogonal to those expected for
simple variations of age and metallicity.

\section{Bi-modal metallicity models}
Our understanding of chemical evolution dictates 
that galaxies, unlike star clusters, have complex chemical make-ups, and so
the comparison of single metallicity SEDs to real galaxies is limited in
its usefulness. Color and Mg$_{2}$ line strength gradients with galactic radius
\citep{DSP93} and 
the width of the RGB in M32 \citep{GLWFFMABHLOS96} are evidence of stellar 
populations of varying metallicities within ellipticals.  
However, the details of the
metallicity distribution in ellipticals are not well understood. Comparing
the Mg$_{2}$ and Fe line strengths for synthetic populations and observed galaxies,
\citet{Greggio96} finds high average 
metallicities, small metallicity dispersions and old ages for ellipticals. 
\citet{WDJ96} find that for M31 and M32
 and \citet{BCF94} find that for
NGC1404 and NGC4649  only $\leq$ 10\% by mass of stars are metal 
poor (Z $\leq$ 0.008), which suggests that the `G dwarf problem' is not
limited to the disk of the Galaxy. 

Because simple, single abundance SEDs did not match the
mid-UV features of the majority of the galaxies in our sample, we looked to
see how a spread in metallicities affects the light of a single 
stellar population.  We consider a very simple bi-modal 
metallicity scenario for old
stellar populations.  Two different bi-modal metallicity models are created
by combining the spectra of the metal-poor Z=0.0004 12 Gyr population 
with the metal-rich spectra
Z=0.05 or Z=0.02 12 Gyr populations.  The models
are constructed so that 0.5\%, 1\%, 5\%, 10\%, and 15\% of the total galaxy
mass originates from the metal-poor 
Z=0.0004 sub-population.  These bi-modal metallicity models are then
measured for Bl2538, Mg$_{2}$, and $S2850$. Figure 13 shows the
combined 10\% Z=0.0004 mass and 90\% Z=0.05 mass 12 Gyr model (solid line)
and each of its sub-population SEDs (short dashed line for Z=0.0004, long 
dashed
line for Z=0.05).  While the metal-rich component dominates the visible
light, the metal-poor component dominates the mid-UV. (Table 5)

In Figures 14a,b,c we show the single metallicity models for a 
given age and
the results of the bi-modal metallicity populations. The Z=0.05 / Z=0.0004
models are represented by the pentagons
where the model with the highest
slope and smallest pentagon has 0.5\% of its mass from the metal-poor 
sub-population and the model with the lowest slope and largest pentagon 
has 15 \% of its mass from the 
metal-poor sub-population.  The Z=0.02 / Z=0.0004 models are represented
by stars, again with 
the smallest star representing the 0.5\% metal-poor mass model and
the largest star representing the 15\% metal-poor mass model. 

Comparing the bi-modal models with
the galaxies, we see that the positions of most of the 
galaxies in the
$S2850$ v. Bl2538 plane
may be explained by a significant contribution
of a metal-poor stellar population to the flux at 2600 {\AA} (Figure 14a).
On the other hand, contamination by ``blue stragglers'' or problems
with the hot component subtraction may also result in weaker mid-UV
features. 
But it is important to
note that a small percentage of metal-poor stars can contribute most of the
light at 2600 {\AA}. While our results do not predict the number
of low metallicity stars needed to ``solve'' the G-dwarf problem, it is
not unreasonable to expect a few low metallicity stars in ellipticals.
However, except for NGC 1399, for which the bi-modal metallicity tracks fit
 particularly well in all three planes, the Mg$_{2}$ strengths are slightly too
strong to agree with most of the galaxies.

\section{Discussion}

The combination of Bl2538, $S2850$, and Mg$_{2}$ provides a unique
tool for investigating elliptical galaxy evolution. The UV features are most
responsive to changes in the effective temperature of the turn-off population, 
while  Mg$_2$ to first order measures the effective temperature of 
the giant branch. The positions of the galaxies in the line-strength
vs. line-strength diagrams suggest: (a) these galaxies are not well 
represented by simple (single-age, single-metallicity) populations, and
(b) that the star-formation and chemical enrichment histories were
unique and complex. The addition of a small metal-poor component to
a dominant old, metal-rich component helps to reconcile the models to
the data.

Part of the difficulty in understanding the observations probably stems
from the heterogeneous nature of the sample of galaxies in our sample,
which to a certain extent represent the extremes of phenomena seen
in elliptical galaxies, rather than the norm. It is thus worth considering
each galaxy in turn, and examining the UV results in the context of
other information about the galaxy.

\emph{M32:} M 32, a compact E1 satellite of M 31, is the closest
example of an elliptical galaxy. M 32 is unusual in that it has a very
weak Mg$_{2}$ feature and no UV upturn. The stellar population in this
galaxy has been the subject of intense study, both via integrated light
and via the resolved stellar population. Based on its blue (B-V) color, 
and on spectral synthesis, \citet{O'Connell80} has suggested that it contains
a $\sim$ 5 Gyr subpopulation. The population synthesis study of \citet{BCF94} found that M 32 spectrum
is dominated by a 13-15 Gyr, Z $\leq$ $\zsun$ population, and has evidence for
an additional 5 Gyr stellar population.  The mid-UV region of M 32's
spectrum is not well fit by the closed box enrichment model and is more 
closely matched 
when the contribution from Z $\leq$ 0.008 stars is removed. In 
\citet{BCF94} and later in \citet{TCBF96}, the infall of gas
is suggested as a mechanism for increasing the galaxy's metallicity quickly
and preventing the formation of a significant number of metal-poor stars.
HST photometry a few arcminutes from the M 32's nucleus 
\citep{GLWFFMABHLOS96} seems to have 
confirmed the dominance of old solar
metallicity stars and the deficiency of metal-poor stars. 
The observed width of the red giant branch requires
that at least half of the stars in M 32 be older than 8 Gyr. Grillmair  et 
al.'s  study also requires that the stars span a wide 
metallicity range but are strongly skewed to solar abundances.

Based on the comparison of IUE mid-UV observations of  M 32's nucleus
with the globular cluster 47 Tuc, \citet{RD99} 
found the mid-UV spectrum of M 32 was best fit by a combination of a
solar metallicity MSTO, a solar giant branch, and a hot A star component
consisting of metal-poor stars and possibly some P-AGB stars.  The
mid-UV spectral breaks 2609/2660 and 2828/2923 and absorption features Mg2800
and Mg2852 are weaker in M 32 than in 47 Tuc, but the C26-30 color (similar
to our $S2850$) is redder in M 32.  While Rose \& Deng (1999) were able to 
reproduce the spectral breaks and absorption lines with a solar MSTO population
with a $\sim$ 25 \% metal-poor contribution of the mid-UV light, the
simulated C26-30 color was $\sim$ 0.1 magnitude bluer than M 32.  Since this
color is dominated by the MSTO whereas the other features are influenced more
by the hotter population, a slightly higher metallicity MSTO may 
reconcile the simulation with M 32.

From the simple single age, single metallicity SEDs (Figures 7-9),
we see the mid-UV and Mg$_{2}$ are not consistent with a single metallicity.
The slope of M32 points to a dominant population with a 
metallicity Z $\geq$ 0.02, the mid-UV Bl2538 value implies a lower 
metallicity population of Z $\sim$ 0.001-0.008, and
 Mg$_{2}$ gives a Z $\sim$ 0.008 - 0.02
population.  Invoking a bi-modal abundance model 
improves the situation slightly. 
The 12 Gyr, 0.5 \% Z=0.0004/ 99.5 \% Z=0.05 SED matches the
UV features, but gives a Mg$_{2}$ value stronger that what is observed (Figure
14a-c).  We have also modeled the mid-UV features for  composite Z=0.05
populations with both a 12 Gyr and 3 Gyr population (Figure 15).  While 
the $S2850$ of M 32 is matched by 5\% 3 Gyr/ 95 \% 12 Gyr Z=0.05 composite
population, the $Bl2538$ of the composite model is too strong.  Solar 
metallicity composite 3 Gyr/12 Gyr  populations would also have $Bl2538$ 
features stronger than M 32 and shallower slopes.  Figure 16 shows a direct
comparison between the M 32 spectra, the best fitting metal-poor component 
model (0.5\% Z=0.0004 / 99.5\% Z=0.05 12 Gyr), and the best fitting young 
component model (5\% 3 Gyr/ 95\% 12 Gyr Z=0.05).  The slopes are identical
for M 32 and the two models, but the $Bl2538$ of the multi-metallicity model
agrees more closely with M 32.
A more complex model with a distribution of metallicities and ages may
reproduce both Mg$_{2}$ and the mid-UV.

\emph{Invoking an intermediate 3-5 Gyr metal-rich
sub-population is not sufficient to explain M 32's UV features; the
presence of a small number of metal-poor stars is implied.} 
The number of low abundance stars predicted is very small ($\sim$ 0.5\%).
 This agrees with 
\citet{BCF94} and \citet{GLWFFMABHLOS96} result of a deficiency of M 32 
metal-poor stars as compared to a closed box chemical evolution model.
However, the existence of a 2-5 Gyr metal-rich population is not ruled out.
The steep slope of M 32's spectrum seems to require a Z $>$ $\zsun$ MSTO 
population (Figures 14a, 15).
It is important to note that our mid-UV M 32 spectrum is of only the
inner-most 80 pc, hence we may be observing the most metal-rich region of 
the galaxy, whereas the HST color magnitude diagram \citep{GLWFFMABHLOS96}
which gives a peak abundance $\sim$ $\zsun$) is of a region $\sim$ 500 pc 
from the nucleus.

\emph{Bulge of M31:}M 31 has one of the reddest and strongest
absorption-featured bulges known.  Its Mg$_{2}$ index in the innermost
regions is comparable to that found for ellipticals and $\alpha$
element enhancement has been suggested \citep{Davidge97}.  However,
its H$\beta$ strength \citep{Davidge97} and the reported presence of
luminous AGB stars suggest an age younger than our Galactic halo of
$\sim$ 8-10 Gyr \citep{RM95}.  Absorption line gradients
suggest that the mean stellar population becomes older and more metal
poor with increasing radius \citep{Davidge97}.

Like M 32, the spectral features can not be fit by a single metallicity model
and are more consistent with an additional low abundance population.
In Figures 7-9, we find that both Mg$_{2}$ and $S2850$ are consistent
with high metallicity (Z $\geq$ 0.02) dominant population, while Bl2538
gives a lower metallicity (Z $\sim$ 0.001).  The 12 Gyr, 0.5 \% Z=0.0004
/ 99.5 \% Z=0.05  model matches M 31 in the mid-UV, but produces a
slightly stronger Mg$_{2}$ than observed (Figure 14 a-c). 
The mid-UV features of the bulge of M 31 
suggest a small number of metal-poor stars ($\leq$ 1\% by mass), 
while its Mg$_{2}$ may point
to the contribution of a metal-rich, intermediate age population. 

\emph{NGC 3610:} This galaxy has many properties of a 3 - 7 Gyr
merger remnant: NGC 3610 has the richest fine structure measured for an
elliptical; it has an unusually blue color; and it possesses a weak
Mg$_{2}$ feature and an enhanced H$\beta$ absorption \citep{SSFBOG90, SS92}.  
There is not a significant amount of dust or cold gas in
the galaxy, nor is there any evidence for ongoing star formation.  HST
observations have detected a number of red globular clusters in NGC
3610, consistent with an age of 4 Gyr \citep{WMSF97}.

Comparison with the single age, single metallicity SEDs shows that NGC
3610's slope and Mg$_{2}$ are consistent with high metallicity (Z
$\geq$ 0.02) dominant population, while the Bl2538 feature implies a
lower metallicity (Z $\sim$ 0.001).  The 12 Gyr, 1\% Z=0.0004 / 99\%
Z=0.05 model is consistent with the mid-UV features, but predicts
a slightly stronger Mg$_{2}$ value than observed in NGC 3610.
 Here again, we detect the
signature of a small number of metal-poor stars ($\leq$ 1\% by mass)
 in the mid-UV light,
and while the Mg$_{2}$ feature may be influenced by a
population of metal-rich 3 -7 Gyr stars.

\emph{NGC 1399:} This gE in  has one of the strongest
Mg$_{2}$ values and UV upturns known.  Like other gEs, NGC 1399 is
thought to be old and very metal-rich.  NGC 1399 possesses $\sim$ 5800
globular clusters; recent Keck spectroscopy \citep{KBSFGH98} of $\sim$ 20 of these clusters give a mean [Fe/H] $\sim$ -0.8 dex
and found no very metal-poor clusters (although the sample may have
been biased against blue clusters).

Surprisingly, both mid-UV features point to a low metallicity
population when compared to the simple models, while the Mg$_{2}$
implies a high metallicity dominant population.  The 12 Gyr 
model with 5-10\% Z=0.0004 / Z=0.05  agrees with the mid-UV features
and Mg$_{2}$ of the IUE data.  The FOS data, which observes only the
center-most region of NGC 1399, imply an even higher metallicity and/or
a lower metallicity population.  Using the \citet{BIdFPT95} calibration
(equation 1), a Z $\sim 2 \times \zsun$, [Mg/Fe] = 0.0, 10-15 Gyr model is
sufficient to explain the mean NGC 1399 Mg$_{2}$ value obtained by the
wide 10 \arcsec $\times$ 20 \arcsec \ aperture (Section 4.2, Table 4).
An $\alpha$ enhanced ([Mg/Fe] $>$ 0.0) model is not required
to explain NGC 1399's strong Mg$_{2}$, but not ruled out.
\emph{NGC 1399 appears to have the
greatest percentage of low abundance stars ($\sim$ 10 \%), 
indicating a
broader metallicity distribution than less massive ellipticals.} 
The strength of Mg$_{2}$ may rule out the existence of an intermediate age
population.

\section{Conclusions}

We have attempted to set additional constraints on the evolution
of elliptical galaxies and spiral bulges by analyzing their
mid-UV spectra. Through comparisons IUE observations of individual
stars to the Kurucz-model fluxes, two spectral indices 
(Bl2538 and $S2850$) are identified that appear to be promising
diagnostics of the main-sequence turn-off population 
(given the current generation of model stellar fluxes).

Single metallicity, single age model populations are constructed
and found to agree well with the mid-UV features of globular clusters.
However most of the of the galaxies do not fall on the single-age, 
single-metallicity tracks. Contamination by blue stragglers may be
partially responsible. Another explanation is that 
the flux from metal-poor stars, invisible in the optical region
of the spectrum, begins to dominate the flux in the mid-UV.
In order to determine the effect of a small percentage (by mass) of
metal-poor stars on the SED of an otherwise metal-rich population,
simple single age, bi-modal metallicity models were created.  For all
the galaxies, a small ($\leq$ 10 \% by mass) number of low metallicity
stars are implied by the mid-UV features.
Metal-poor stars are of course expected from chemical-evolution
arguments, but evidence for a metallicity spread has been hard
to find in the optical spectra. The metallicity distribution 
function can no longer be ignored if we hope match simultaneously the
mid-UV and the optical spectra.

It is difficult to
draw strong conclusions about the recent star-formation history of the
galaxies in our sample from the mid-UV.  NGC 1399, one of the
strongest Mg$_{2}$ and UV upturn galaxies known, is consistent with a
multi metallicity 12 Gyr model in both the mid-UV and Mg$_{2}$
strength and may possess the greatest percentage of low metallicity stars
observed.  The strength of Mg$_{2}$ suggests that NGC 1399 has few, if
any, stars younger than 10 Gyr. However, for the remaining galaxies, 
especially M 32,  a more
sophisticated model invoking both  old metal-poor and intermediate age 
metal-rich stars may be needed to reconcile the Mg$_{2}$ and mid-UV features.
On the other hand, variations the blue straggler populations and $\alpha$ 
element abundance could also account for the disparities between the optical
and mid-UV spectra.

Because they probe the main-sequence turn-off population, 
mid-UV spectral diagnostics can in principle provide a a powerful 
diagnostic of the star-formation history and chemical evolution of
elliptical galaxies. Our study presents a rather discouraging,
but nevertheless realistic, picture of the current status of the 
models and the data in this wavelength regime. The mid-UV features
we have identified suggest that significant progress can be 
made with additional high-quality spectra of elliptical galaxies,
even if the model stellar fluxes do not improve. However, the 
promise of this wavelength regime for untangling long-standing
issues of galaxy evolution should also motivate attempts to 
improve the model stellar fluxes, which will ultimately put such 
analyses on firmer ground, and allow the use of additional spectral
features.

We wish to thank the referee for his many constructive comments and 
suggestions. We also thank Daniela Calzetti and Anne Kinney for allowing us 
to use their extracted IUE and optical spectra, and Michael Gregg for 
useful discussions about M 32.  This paper is based in part on observations
with the NASA/ESA Hubble Space Telescope obtained at the Space Telescope
Science Institute. Support for this work was provided by NASA through grant 
number GO-03647.01-91A from the Space Telescope Science Institute, which is
operated by the Association of Universities for Research in Astronomy, 
Incorporated, under NASA contract NAS5-26555.

\clearpage
\begin{deluxetable}{ccc}
\footnotesize
\tablecaption{Mid-UV\tablenotemark{a} \  and Optical\tablenotemark{b} \ 
 Spectral Indices}
\tablewidth{0pt}
\tablehead{
\colhead{Index} & \colhead{Absorption Band ({\AA} )} & \colhead{Continuum Bands
({\AA} )} 
}
\startdata
Fe2402  &2382-2422 &2285-2325  \\
& &2432-2458 \\
Bl2538  &2520-2556 &2432-2458 \\
&  &2562-2588 \\
Fe2609  &2596-2622 &2562-2588 \\
& &2647-2673 \\
Mg2800 &2784-2814&2762-2782 \\
&  &2818-2838 \\
Mg2852  &2839-2865 &2818-2838 \\
& &2906-2936 \\
MgWide  &2670-2870 &2470-2670 \\
&  &2930-3130 \\
Fe3000  &2965-3025 &2906-2936 \\
&  &3031-3051 \\ 
H$\beta$   &4848-4877&4828-4848 \\
&  &4877-4892 \\ 
Mg$_{2}$  &5154-5197 &4895-4958 \\
&  &5301-5366 \\
\enddata
\tablenotetext{a} {Fanelli et al. 1990}
\tablenotetext{b} {Faber et al. 1985}
\end{deluxetable}

\begin{deluxetable}{cccccccc}
\footnotesize
\tablecaption{Globular Cluster Data\tablenotemark{a}}
\tablewidth{0pt}
\tablehead{
\colhead{Cluster\tablenotemark{b}} & \colhead{Z} & \colhead{EW} & \colhead{$C_{2466}$} & \colhead{$C_{2959}$} & \colhead{$C_{3122}$} & \colhead{Bl2538} &\colhead{$S2850$}\\
\colhead{} & \colhead{} & \colhead{{\AA} } & \colhead{$\frac{log(F)}{log(F_{2646})}$} 
& \colhead{$\frac{log (F)}{log(F_{2646})}$} & \colhead{$\frac{log(F)}{log(F_{2646})}$}
& \colhead{mag} & \colhead{$\frac{\Delta mag }{1000 {\AA} }$}
}
\startdata
G2b &0.006  &28.6 &0.53 &1.55 &2.07 &0.593  &-1.65 \\
G47Tuc &0.004 &34.2 &0.46 &1.46 &2.06 &0.759 &-1.60 \\
G3b &0.0011 &20.7 &0.63 &1.27 &1.33 &0.396  &-0.68 \\
G3r &0.0010 &23.4 &0.66 &1.24 &1.42 &0.496  &-0.80 \\
G4b &0.0007 &12.7 &0.80 &1.12 &1.24 &0.224  &-0.48 \\
G4r &0.0006 &21.6 &0.68 &1.23 &1.30 &0.416  &-0.63 \\
G5  &0.0002 &9.1 &0.85 &1.18 &1.27 &0.156  &-0.55 \\
\enddata
\tablenotetext{a} {EW and continuum points $C_{2466}$, $C_{2959}$, $C_{3122}$
 were taken from Bonatto et al. 1995}
\tablenotetext{b} {Names of composite globular cluster spectra; r = red, b = blue}
\end{deluxetable}

\begin{deluxetable}{ccc}
\label{tableFOS}
\footnotesize
\tablecaption{FOS Observations}
\tablewidth{0pt}
\tablehead{
\colhead{Galaxy} & \colhead{Grating} & \colhead{Exposure Time}
}
\startdata
NGC1399 &	G130H &	10453s \\
 &	G190H &	5284 \\
 &	G270H &	3542 \\
 &           & \\
NGC3610 &	G130H &	10453 \\
 &	G190H &	5836 \\
 &	G270H &	3531 \\
\enddata
\end{deluxetable}

\begin{deluxetable}{cccccc}
\footnotesize
\tablecaption{Galaxy Data\tablenotemark{a}}
\tablewidth{0pt}
\tablehead{\colhead{Galaxy} & \colhead{Bl2538} & \colhead{Bl2538}& \colhead{$S2850$} & \colhead{$S2850$} & \colhead{Mg$_2$}\\
\colhead{} &\colhead{} & \colhead{corrected} & \colhead{} & \colhead{corrected}
& \colhead{}
}
\startdata
M 32  &0.45 &-  &-2.92 &-  &0.194  \\
M 31   &0.34 &0.49 &-2.29  &-2.73   &0.286   \\
NGC 1399  &0.07 &0.30 &-1.02  &-1.65   &0.312  \\
NGC 3610\tablenotemark{b} &0.30 &0.38 &-2.59 &-2.73  &0.271\tablenotemark{c} \\
NGC 1399\tablenotemark{b} &0.08 &0.23 &-0.80  &-2.04  &0.357\tablenotemark{d}\\
\enddata
\tablenotetext{a} {UV data extracted from IUE database by D. Calzetti;
optical data from Kinney et al. 1996}
\tablenotetext{b} {FOS data, see section 4}
\tablenotetext{c} {Worthey et al. 1992}
\tablenotetext{d} {Mg$_2$ value of nucleus from Burstein et al. 1988}
\end{deluxetable}

\begin{deluxetable}{ccc}
\footnotesize
\tablecaption{12 Gyr Bimodal metallicity models}
\tablewidth{0pt}
\tablehead{\colhead{\% Z=0.0004 mass} & \colhead{Z=0.0004 / Z=0.02 model} & \colhead{Z=0.0004/ Z=0.05 model} \\
\colhead{} & \colhead{\% Z=0.0004 mid-UV flux} & \colhead{\% Z=0.0004 mid-UV flux}}
\startdata
0.5\%  &11\% &35\%  \\
 1\%  &19\% &52\% \\
 5\%  &55\% &85\% \\
10\%  &72\% &92\% \\
15\%  &81\% &95\% \\
\enddata
\end{deluxetable}

\clearpage
\onecolumn
\begin{figure}
\plotone{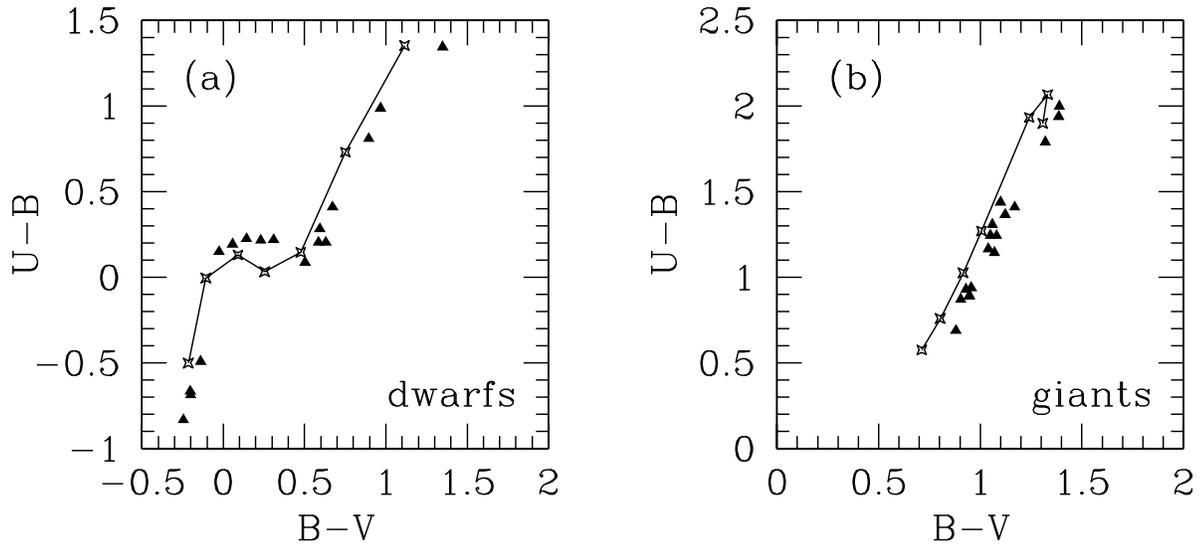}
\caption{(a,b)  $(U-B)$ v. $(B-V)$ for Kurucz solar metallicity
dwarf (a) and giant (b) stellar atmospheres (crosses connected by lines)
against the Gunn-Stryker stars (filled triangles).  
}
\end{figure}

\clearpage
\begin{figure}
\plotone{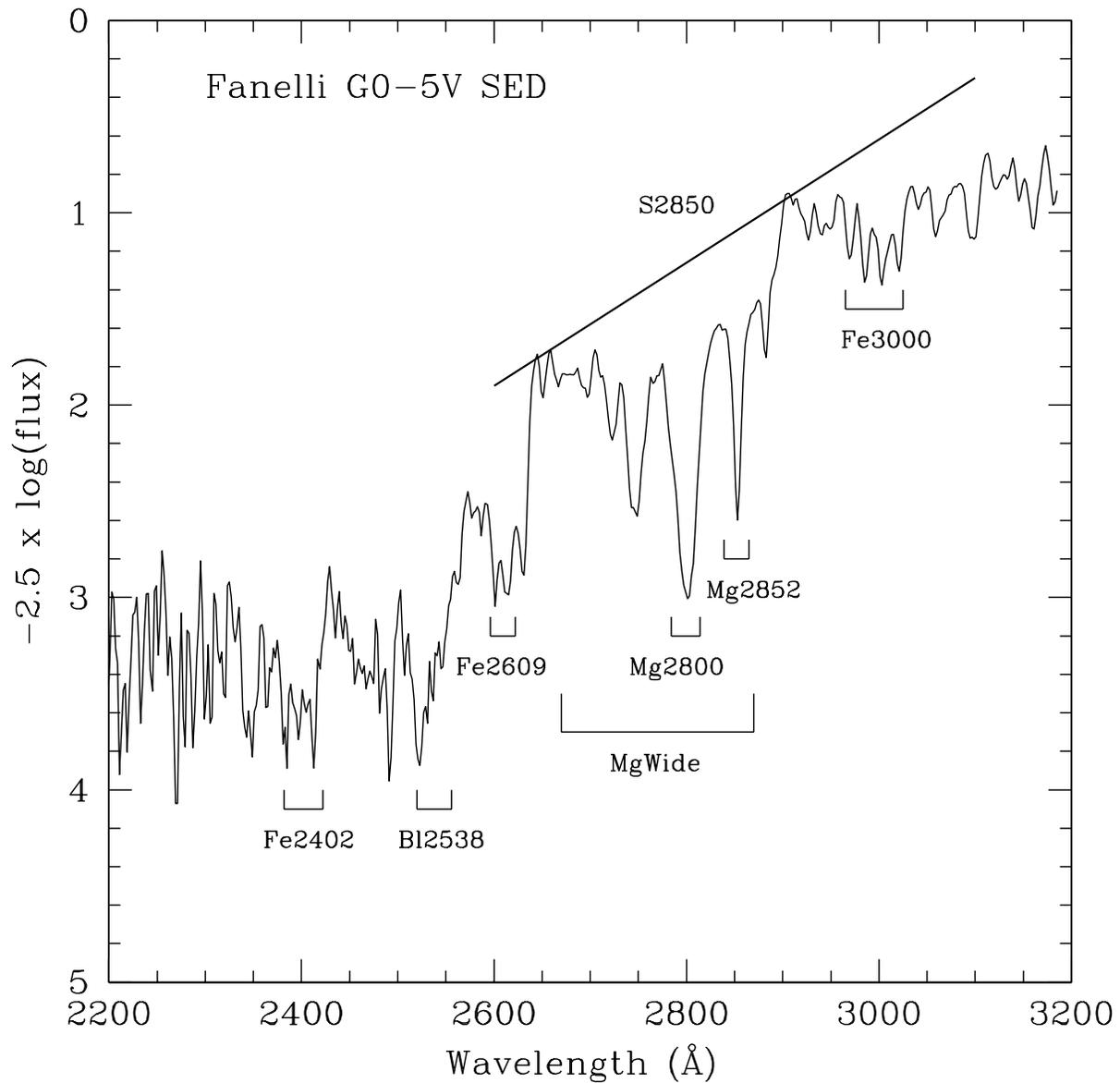}
\caption[f2.ps]{
The Fanelli composite IUE spectra for G0-5V stars is shown with the 
mid-UV spectral indices and $S2850$ identified (Table 1).  $S2850$ is
defined as the continuum slope between 2600 {\AA} and 3100 {\AA}, excluding
the absorption features.
}
\end{figure}

\clearpage
\begin{figure}
\plotone{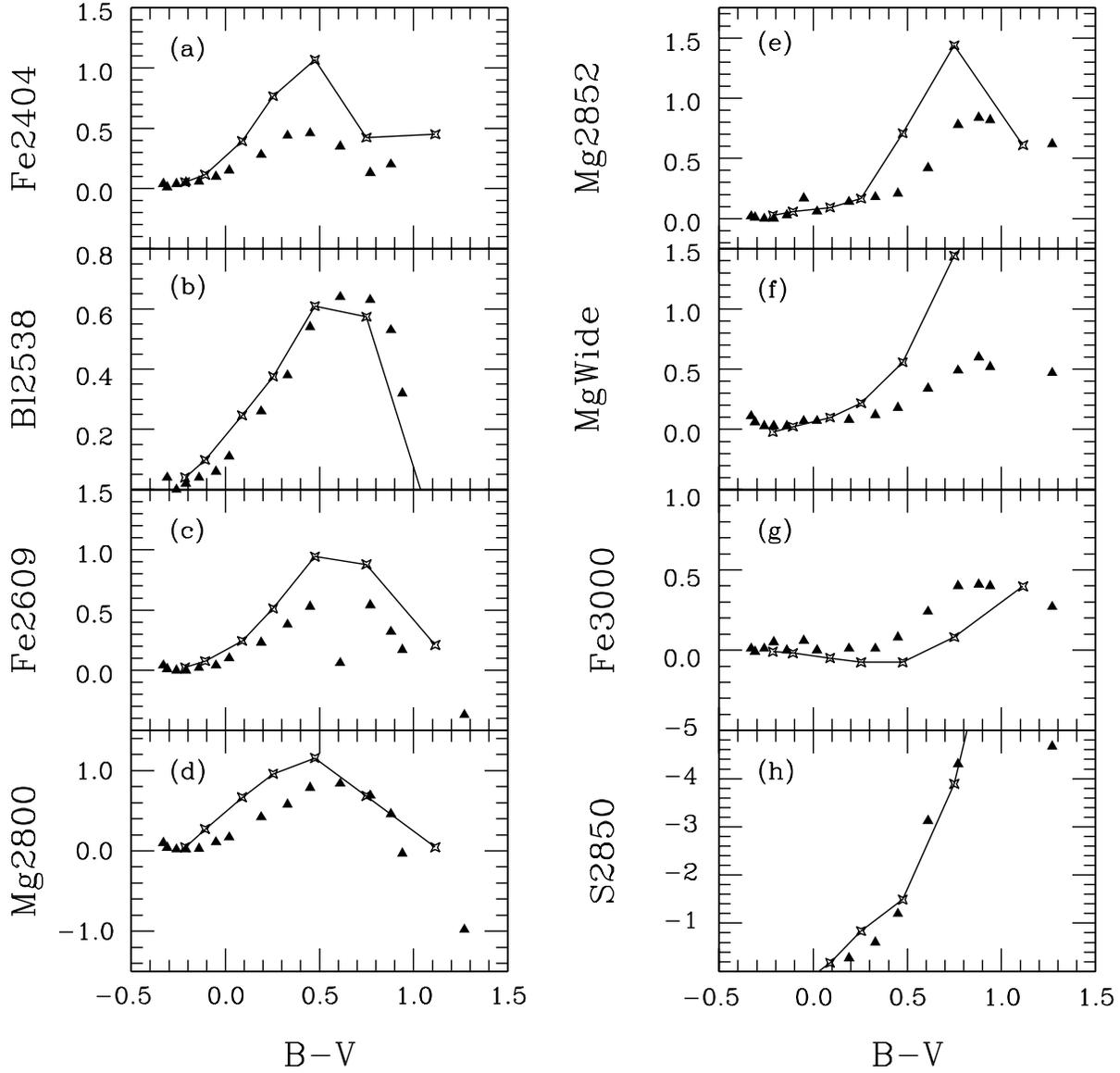}
\caption{(a-h)
The mid-UV absorption lines and continuum values for Kurucz model Z=0.02
dwarfs (crosses connected by lines) are plotted against $(B-V)$ 
with observed values for Fanelli
IUE composite dwarf spectra (triangles).  Note that only the Bl2538 (b) and 
$S2850$ (h) are well modeled by the Kurucz atmospheres.
}
\end{figure}

\clearpage
\begin{figure}
\plotone{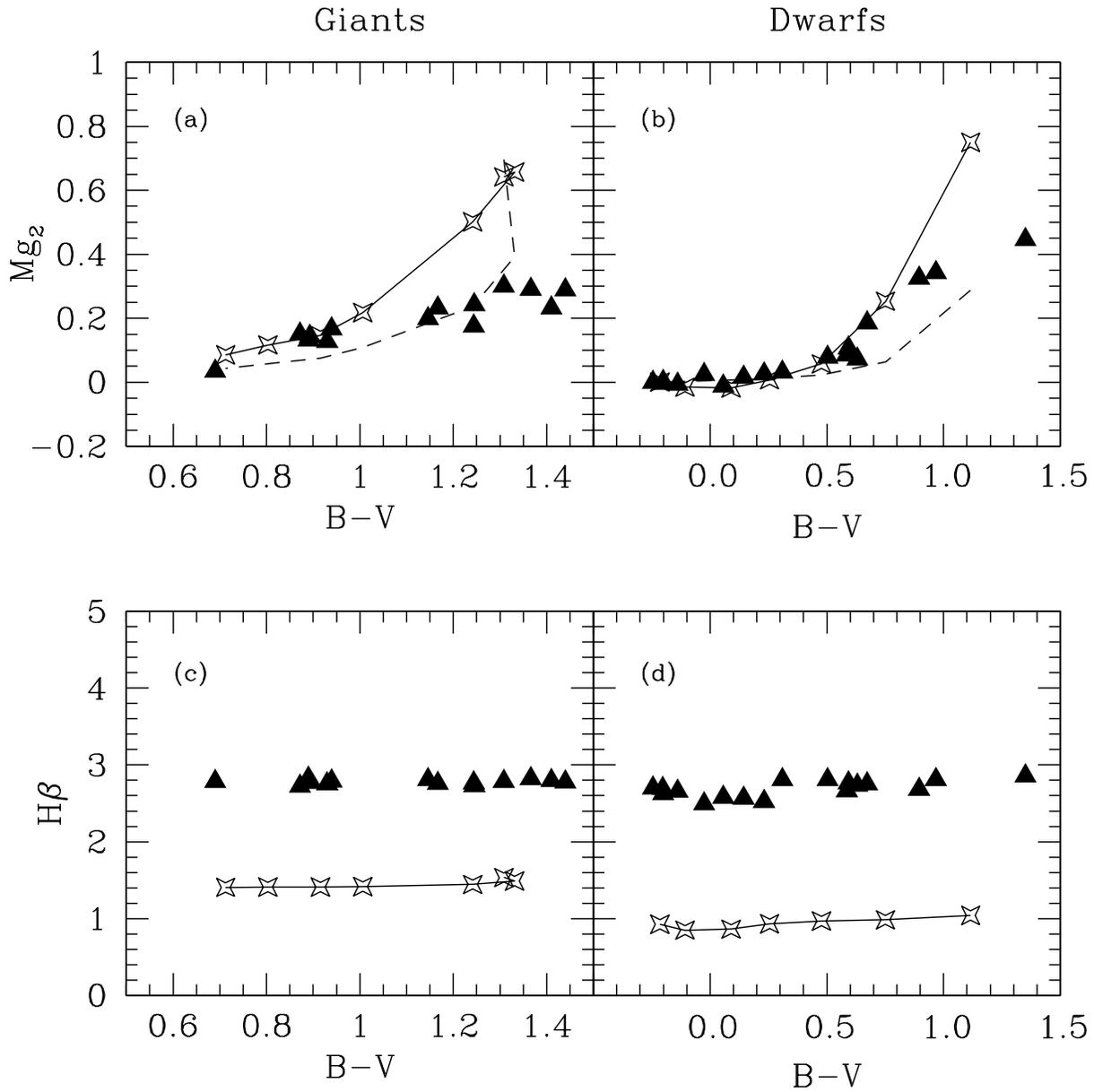}
\caption{
(a,b)The optical absorption feature Mg$_{2}$  for the Kurucz model
Z=0.02 dwarfs and giants
 (crosses connected by lines) are plotted against $(B-V)$ with observed
values for dwarfs and giants taken from the Gunn-Stryker catalog (triangles).
(c,d) Same for H$\beta$.
 }
\end{figure}

\clearpage
\begin{figure}
\plotone{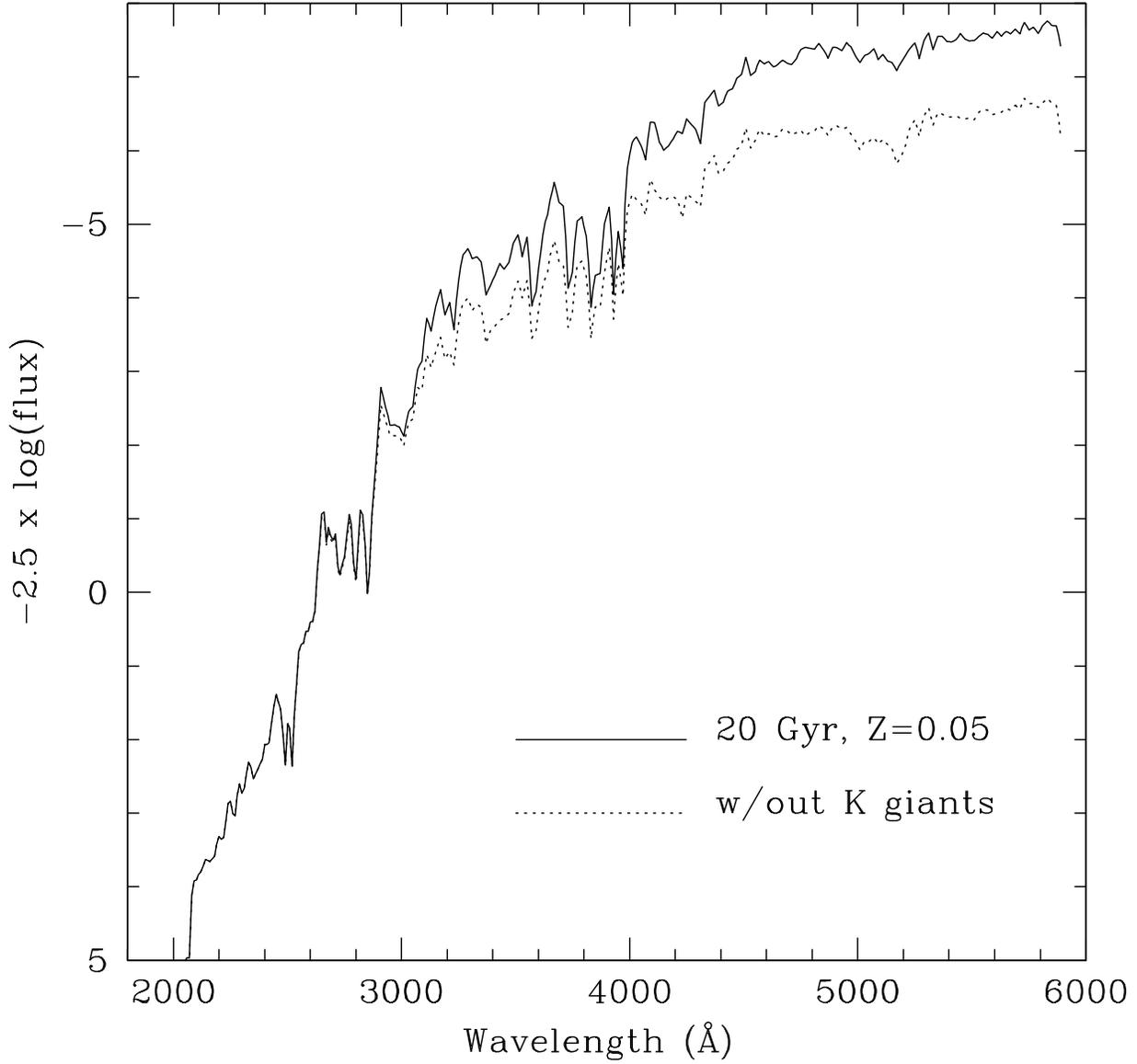}
\caption{
20 Gyr, Z=0.05 SED created with cool giants (solid line) and without cool 
giants (dotted line).  Note that the cool giants 
contribute very little light to 1800 - 3200 {\AA} but 
do dominate the Mg$_{2}$ at $\sim$ 5170 {\AA}.
}
\end{figure}

\clearpage
\begin{figure}
\plotone{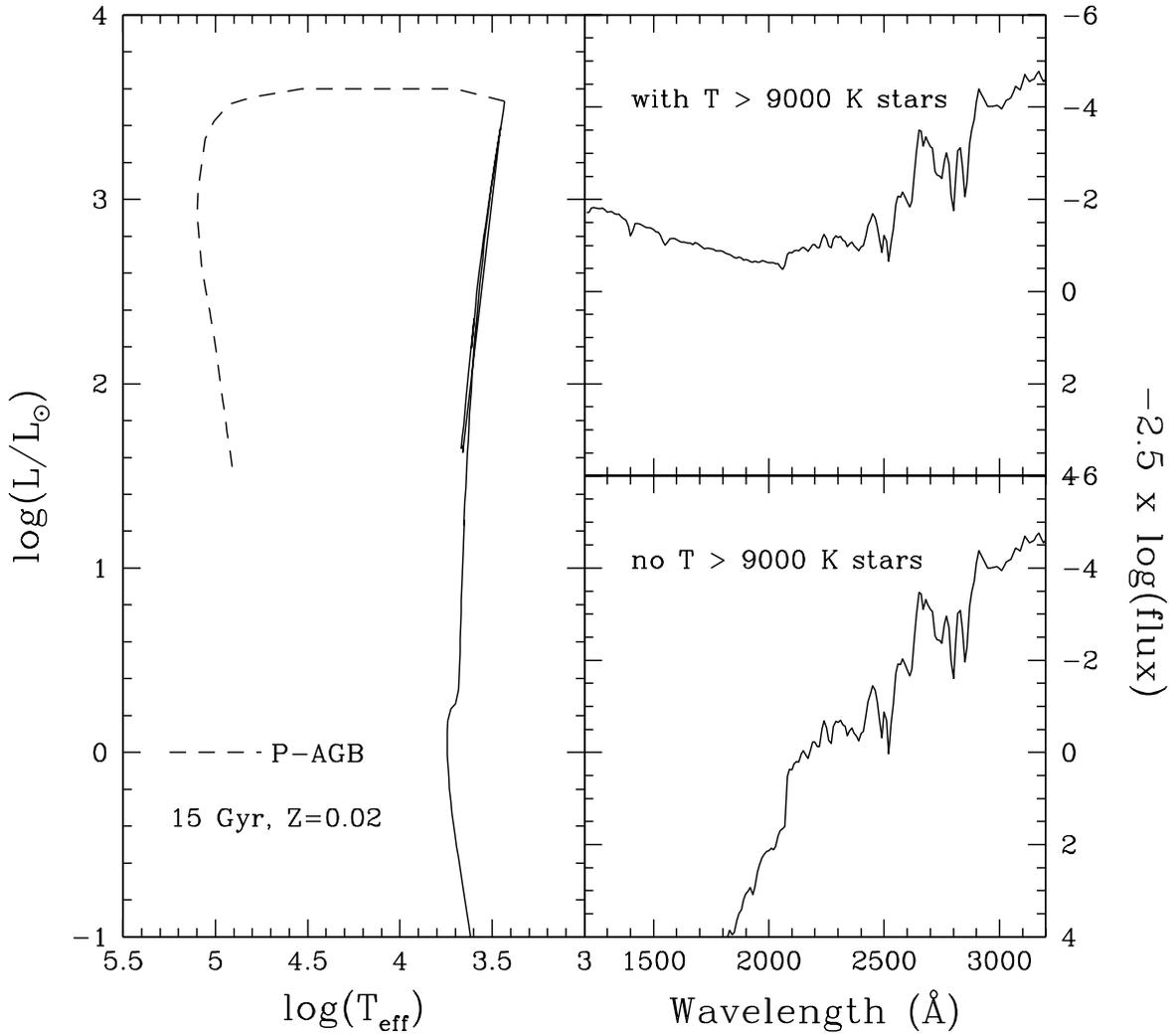}
\caption{
The left hand panel show the Padova isochrone for a 15 Gyr, solar population.
The dashed line is the P-AGB.  The right-hand panels show the 
UV region for the SED (Wavelength v. -2.5 $\times$ log(flux)) 
calculated with and without stars hotter than 9000 K. Note that the
far-UV (wavelength $\geq$ 2000 {\AA} ) is dominated by flux from the P-AGB.
}
\end{figure}

\clearpage
\begin{figure}
\plotone{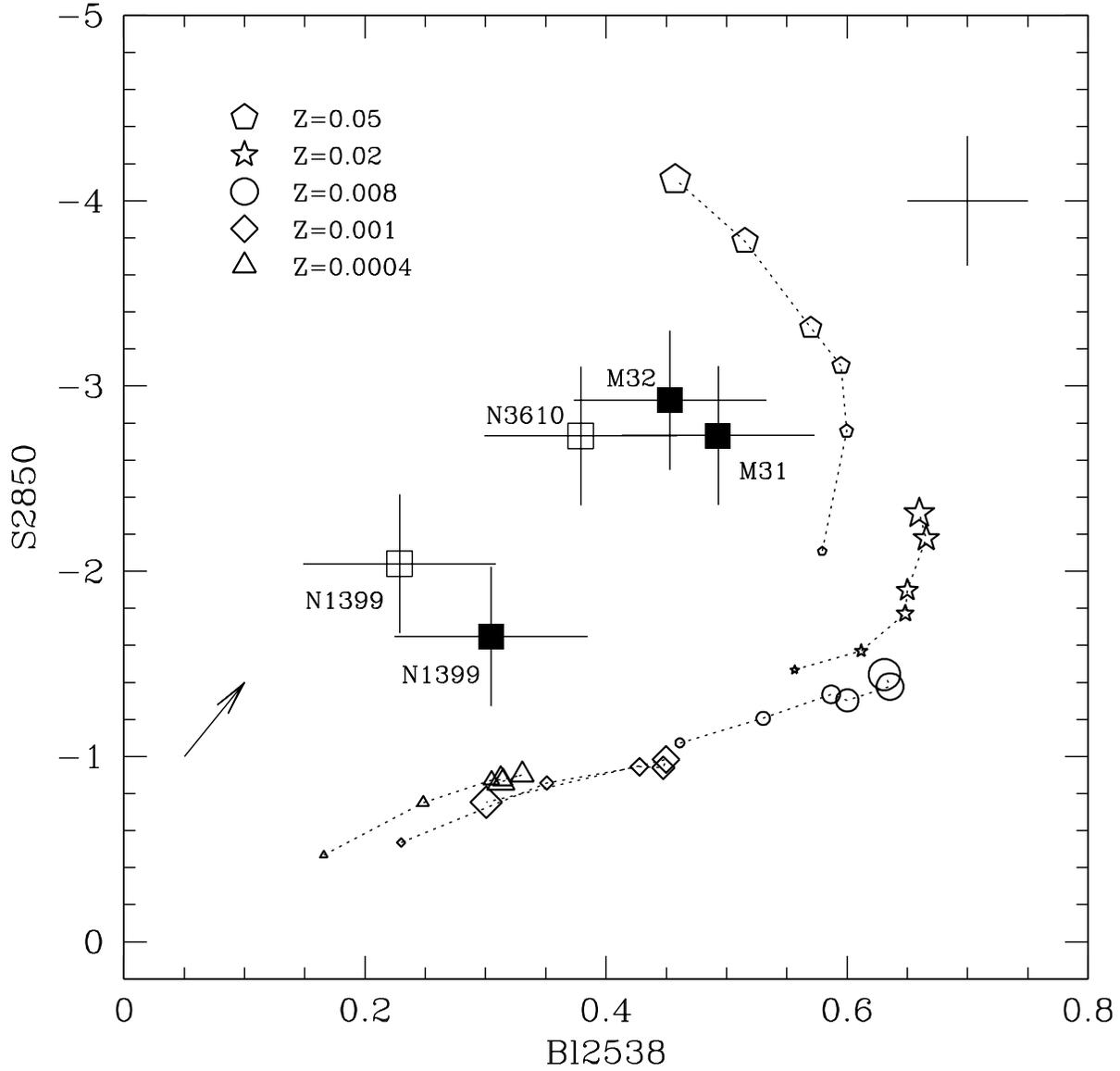}
\caption{
Bl2538 index v. $S2850$ for the single age
(3, 7, 10, 12, 15, and 20 Gyr), single metallicity (Z=0.0004, 0.001, 0.008,
0.02, and 0.05) SEDs with {\it no hot stellar component}. 
Different metallicities have different symbols defined
by the legend
and increasing age models are represented by increasingly larger symbols 
with the smallest symbol = 3 Gyr and the largest symbol = 20 Gyr.
The galaxies are plotted as squares (open for FOS, closed for IUE).
The uncertainty for the model SEDs is given by the error bar in the upper 
right corner. The effect of reddening for A$_B$ = 0.5 is shown by the arrow
in the lower left.
}
\end{figure}

\clearpage
\begin{figure}
\plotone{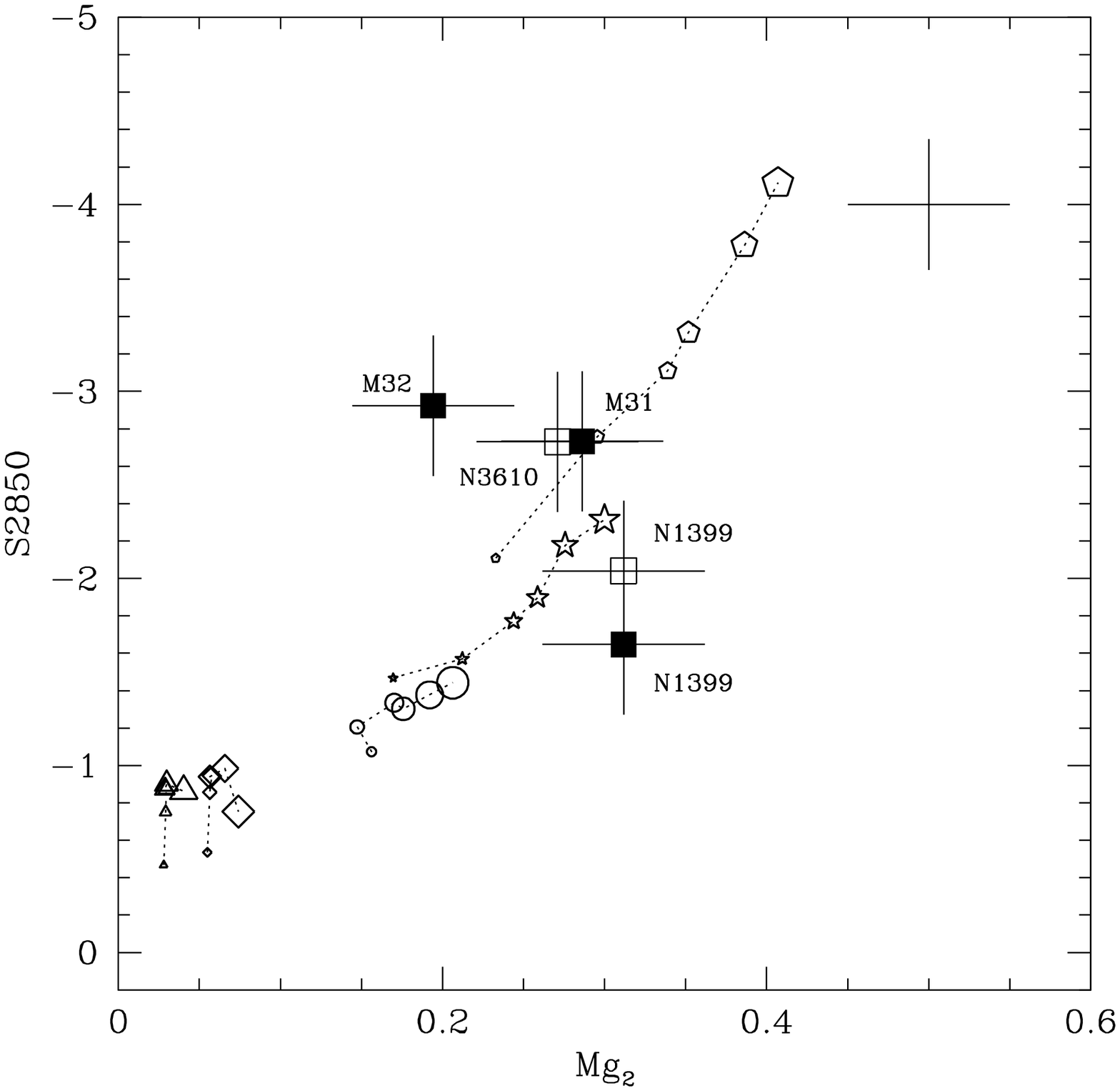}
\caption{
$S2850$ v. Mg$_{2}$ for single age, single metallicity models
with {\it no hot stellar component} . Symbols are
the same as for Figure 7.
}
\end{figure}

\clearpage
\begin{figure}
\plotone{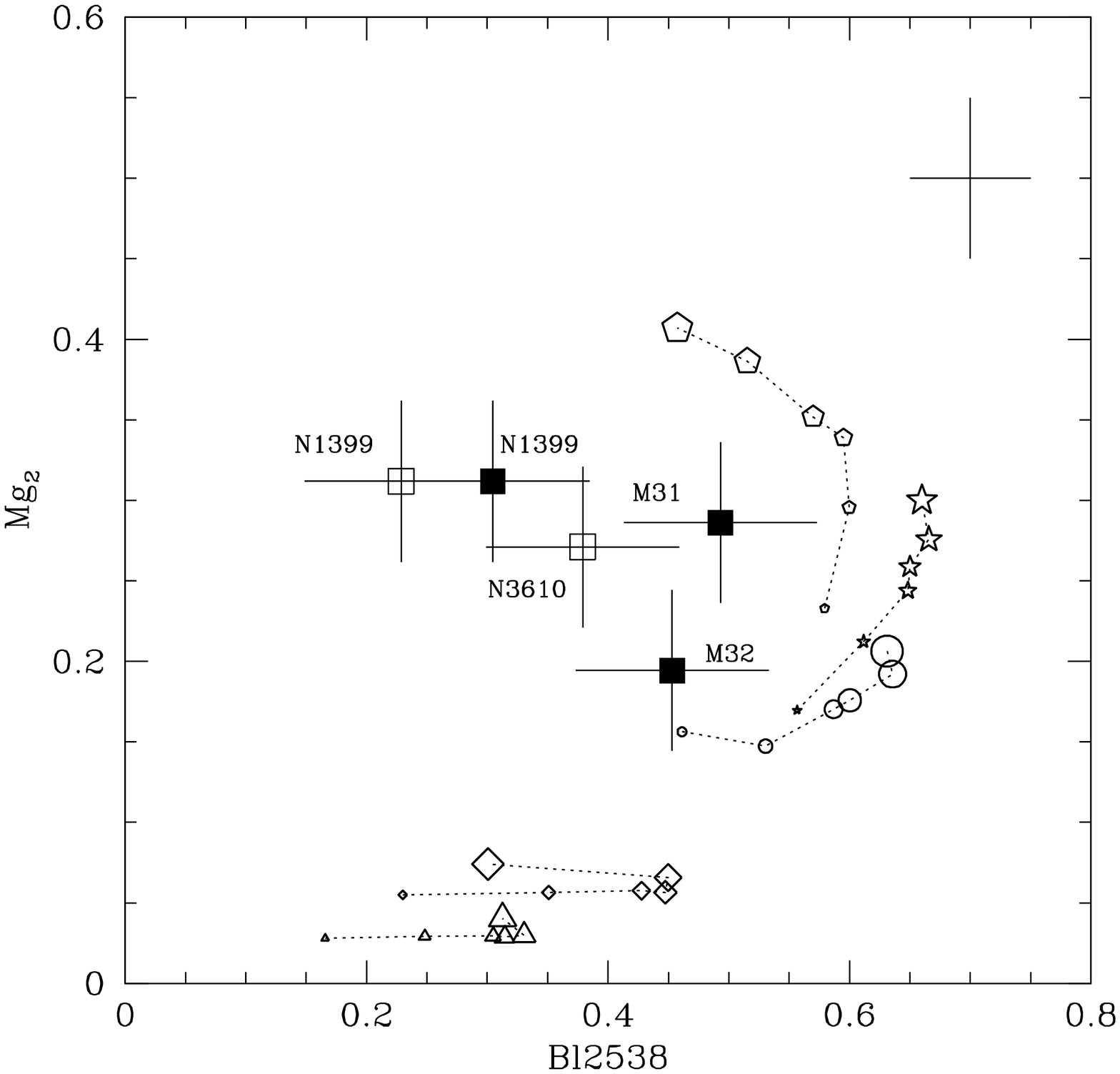}
\caption{
Bl2538 v. Mg$_{2}$ for single age, single metallicity models
with {\it no hot stellar component}. Symbols are
the same as for Figure 7. 
}
\end{figure}

\clearpage
\begin{figure}
\plotone{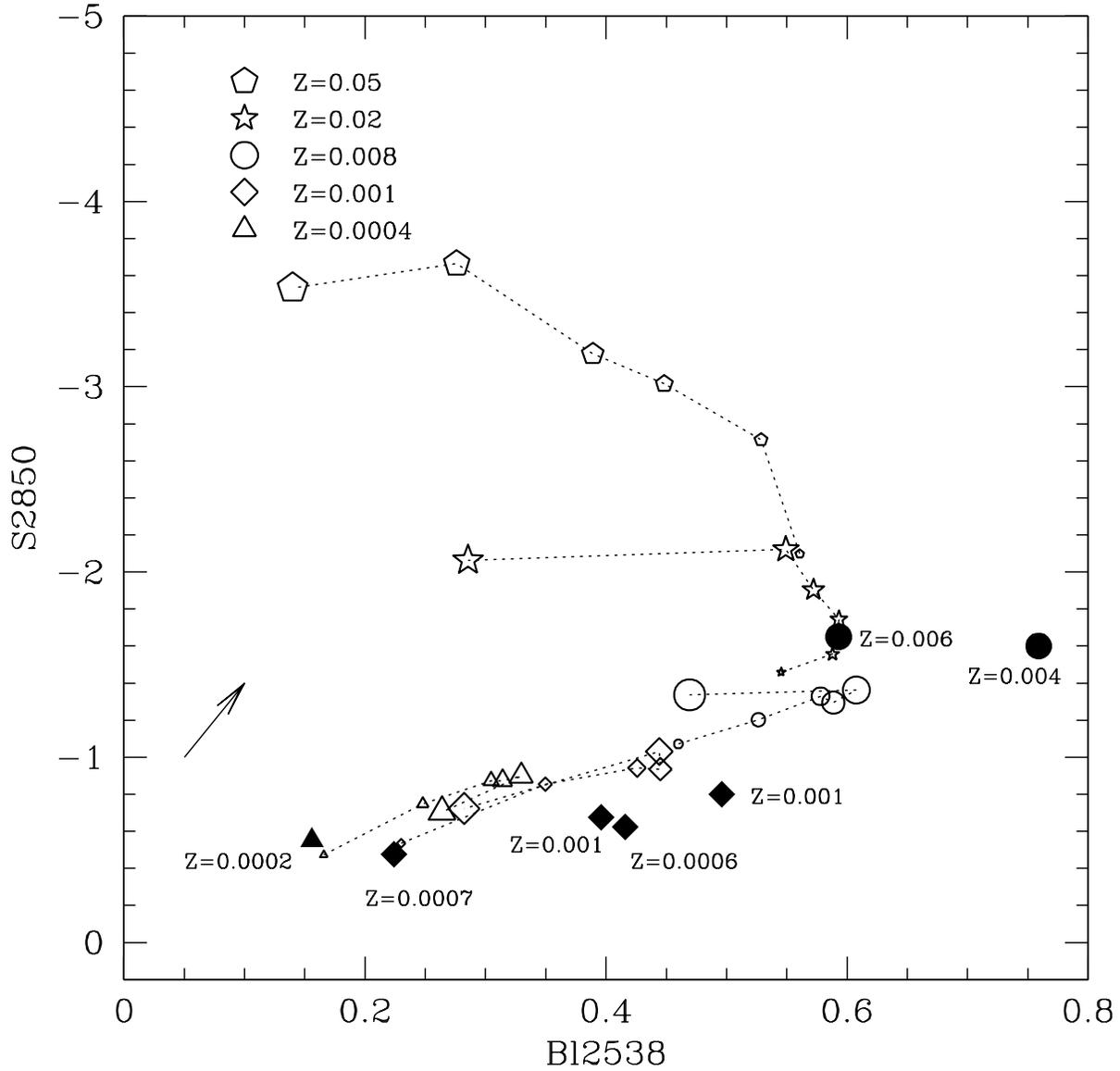}
\caption{
Bl2538 v. S2850 for single age, single metallicity models computed
with the original isochrone and {\it including the hot HB and post HB stars}.
The composite IUE globular cluster values (Table 2, Bonatto et al 1995) 
are plotted as filled symbols, with the shape of the symbol corresponding to
cluster metallicity.  
}
\end{figure}

\clearpage
\begin{figure}
\plotone{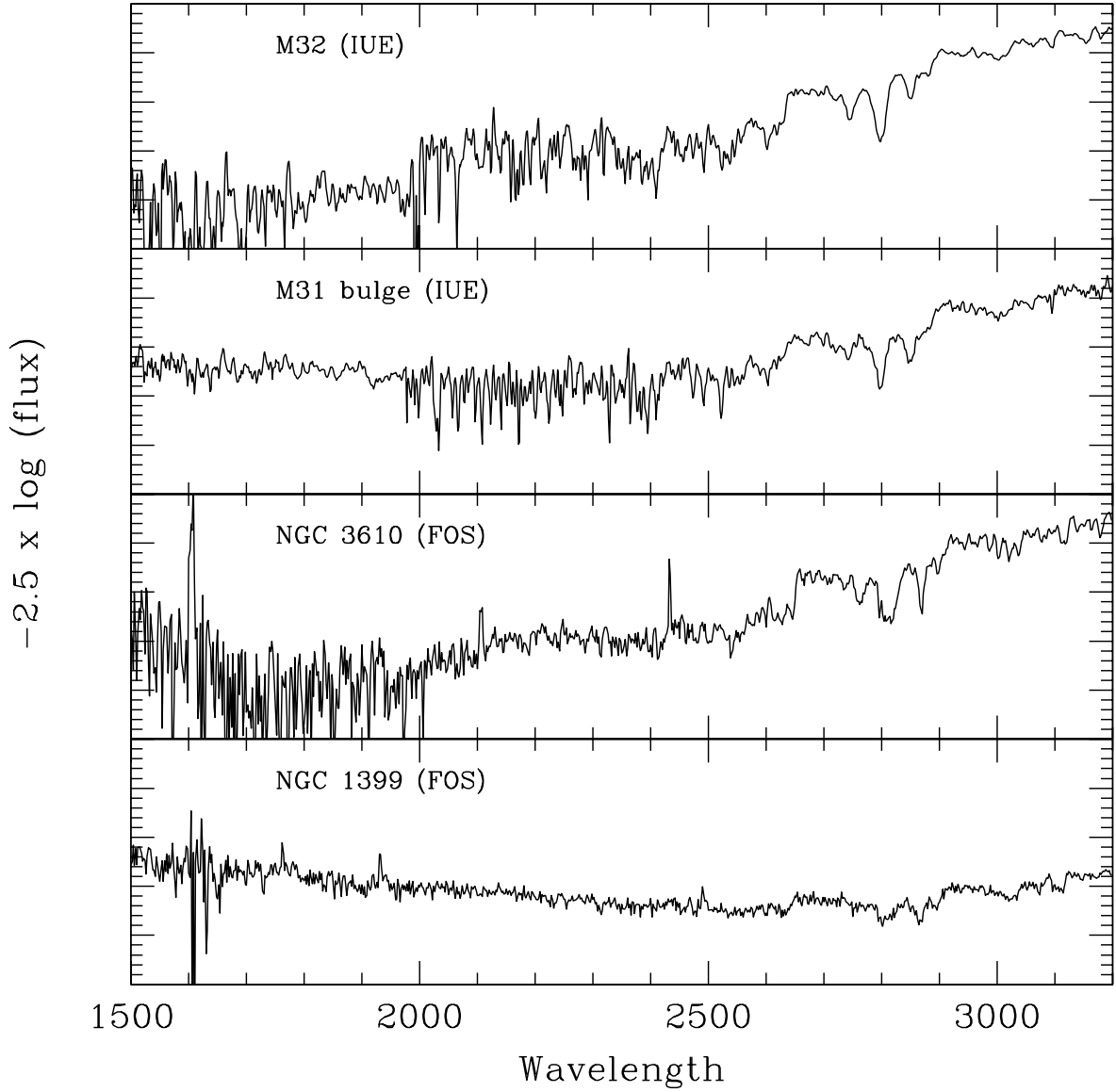}
\caption{
The mid-UV spectra of M32, the bulge of M31, NGC 3610, and NGC 1399.  The FOS
data has been binned to 2 {\AA} resolution.  
Each large tickmark corresponds to 1 magnitude.
}
\end{figure}

\clearpage
\begin{figure}
\plotone{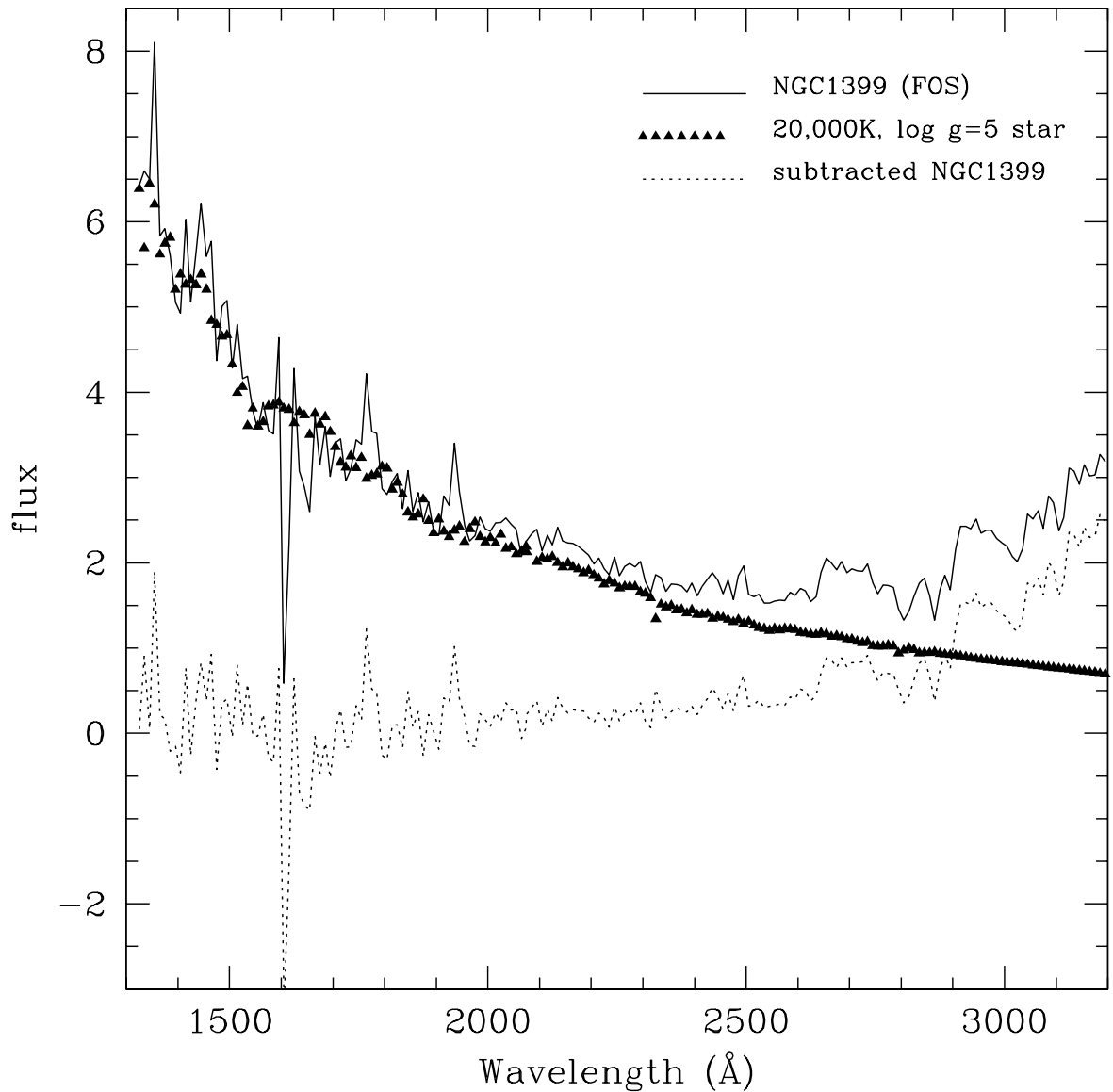}
\caption{
The FOS spectra of NGC 1399 is plotted (solid line) with the 20,000 K, log g
= 5.0 Kurucz atmosphere fitted to NGC 1399 far UV flux (triangles).  
The far UV subtracted
NGC 1399 spectra is given by the dashed line.
}
\end{figure}

\clearpage
\begin{figure}
\plotone{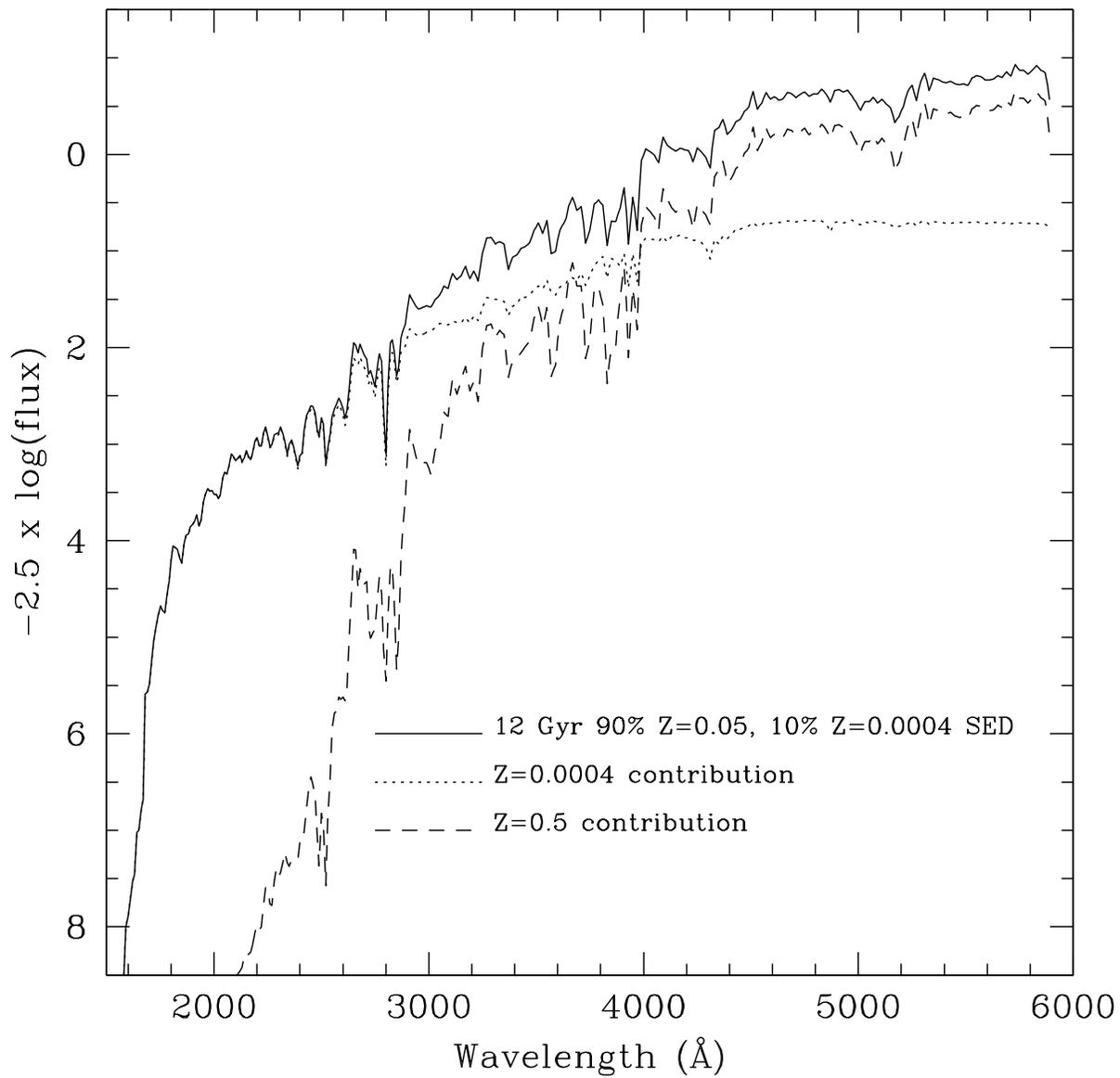}
\caption{
Bi-modal metallicity 12 Gyr SED (solid line) 
where 10 \% of mass is from a Z=0.0004 
population (dot-dashed line) and 90 \% from a Z=0.05 population (dashed line).
}
\end{figure}

\clearpage
\begin{figure}
\plotone{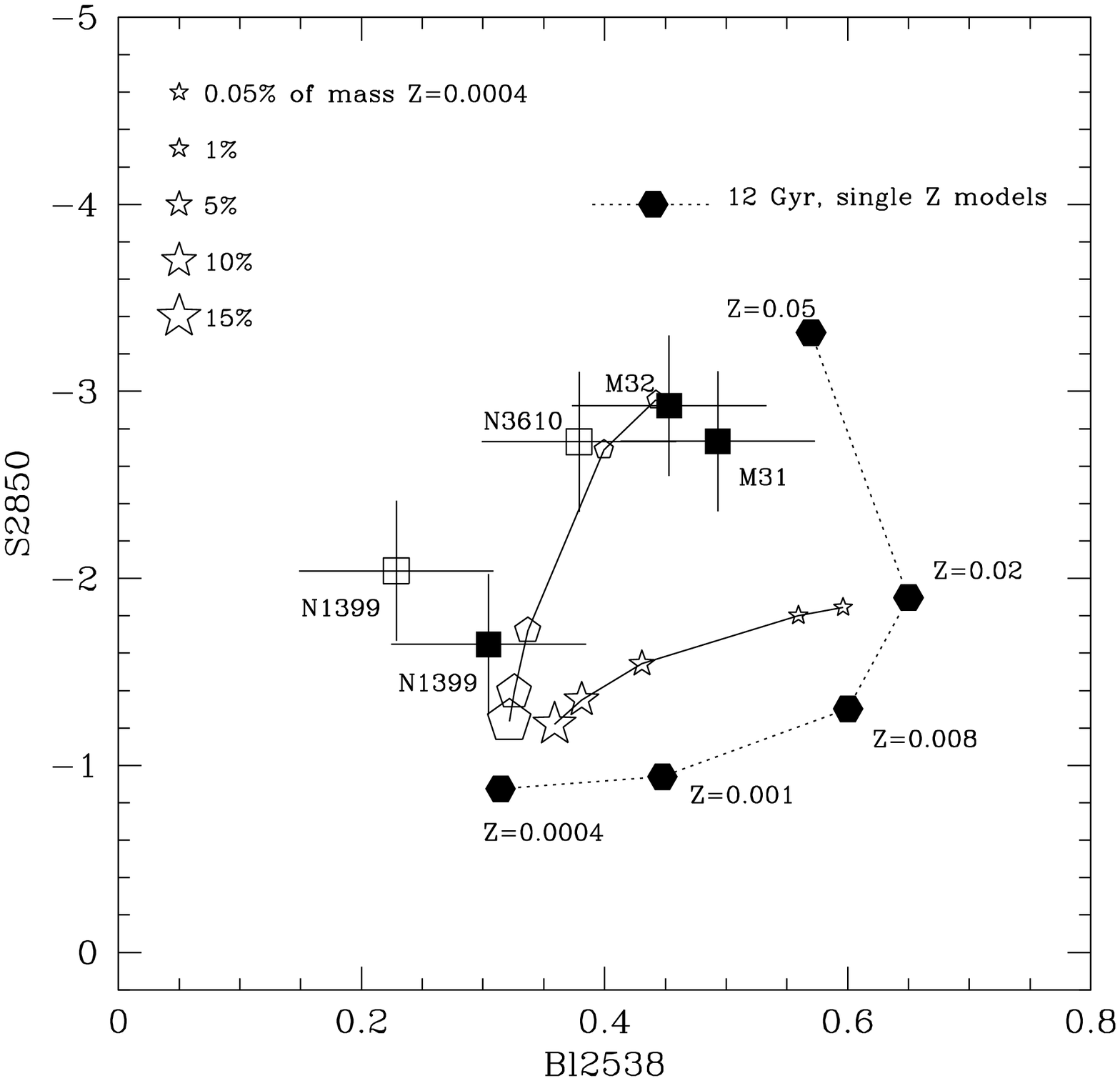}
\end{figure}

\clearpage
\begin{figure}
\plotone{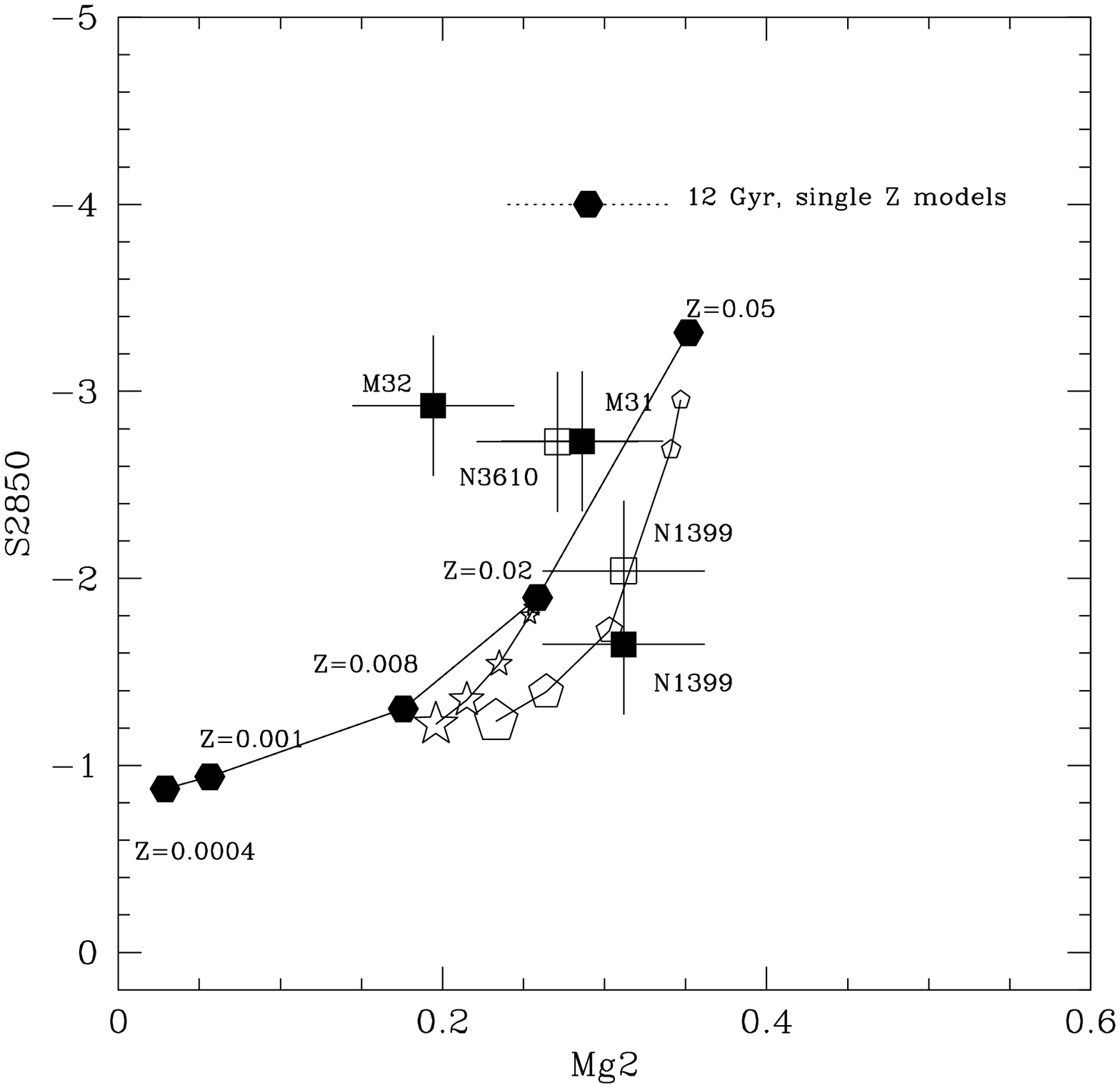}
\end{figure}

\clearpage
\begin{figure}
\plotone{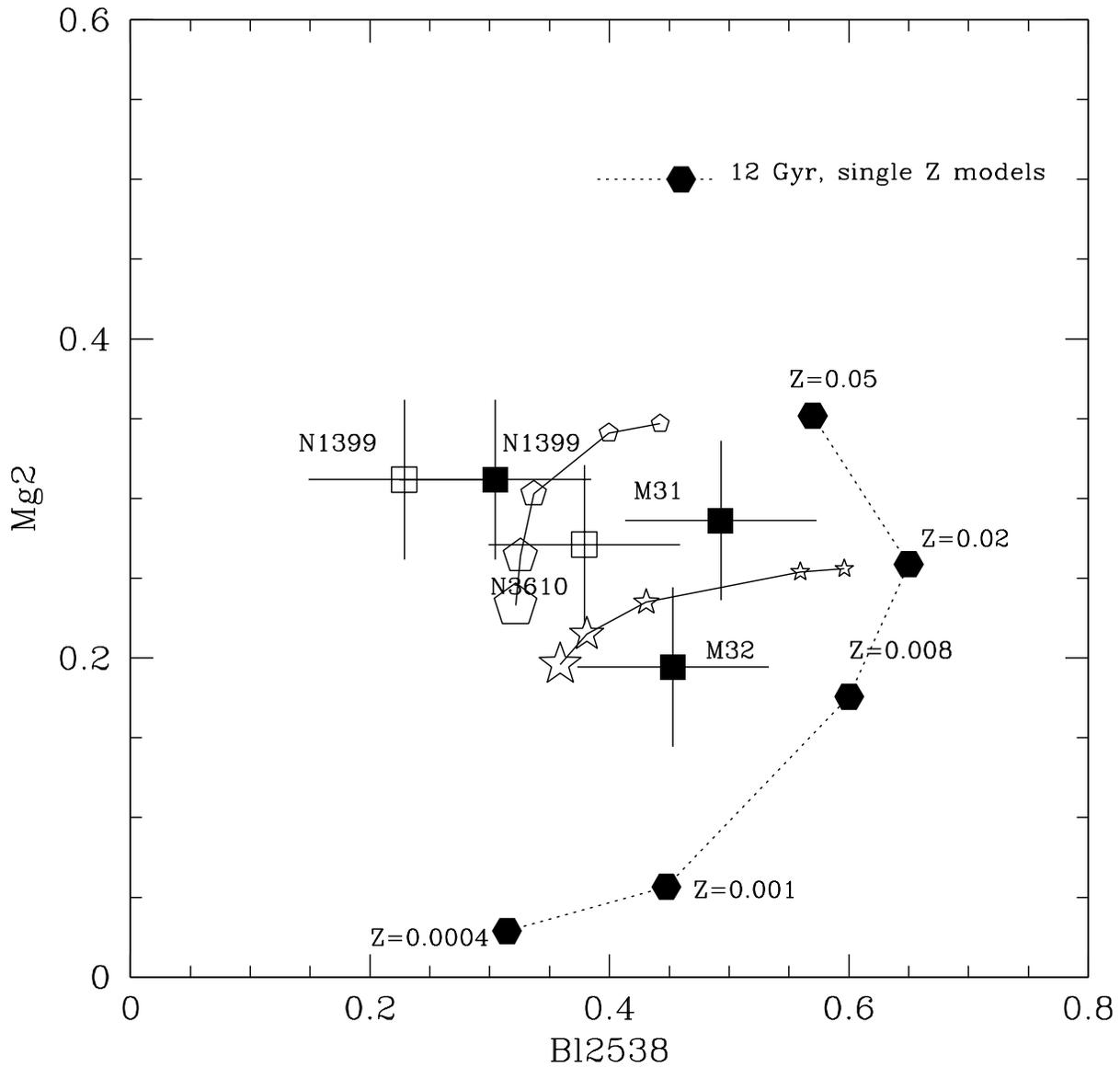}
\caption{(a-c) $S2850$ v Bl2538, $S2850$ v Mg$_{2}$, Mg$_{2}$ v 
Bl2538;
Bi-modal metallicity models for 12Gyr,  Z=0.05/0.0004
models are represented by the pentagons, and the Z=0.02/0.0004 models are
represented by the stars. The models given have 0.5\%, 1\%, 5\%,
10\% and 15\% of total mass from the metal-poor subpopulation.
For comparison, the 12 Gyr single abundance models are given as well.
}
\end{figure}

\clearpage
\begin{figure}
\plotone{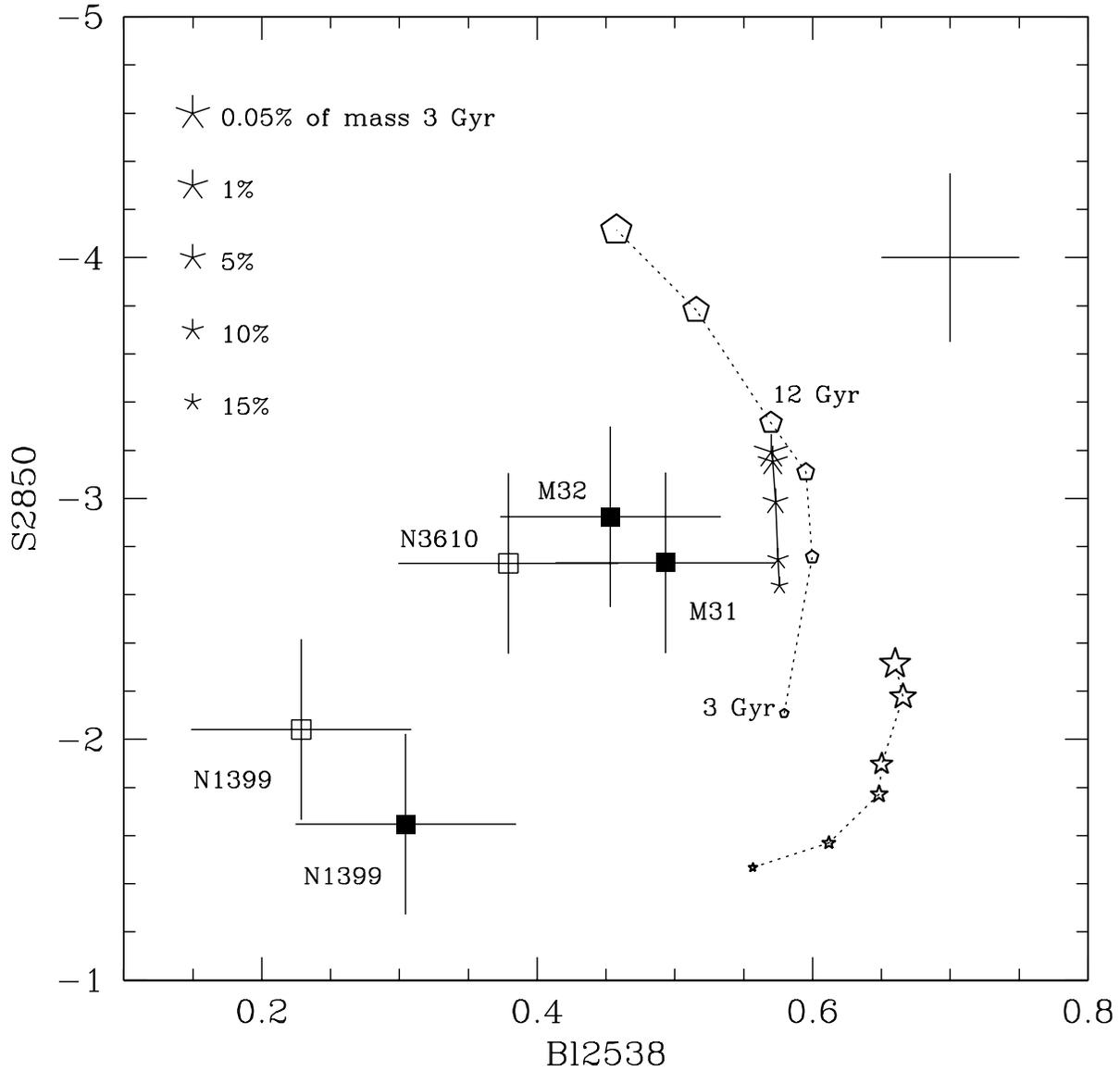}
\caption{Composite 3 Gyr/12 Gyr Z=0.05 models (asterisks) are computed
for increasing mass of the young 3 Gyr sub-population.  The single age Z=0.05
(pentagons) and Z=0.02 (stars) are shown as well for comparison.
Note that M 32 is not matched by any of the composite models.}
\end{figure}

\clearpage
\begin{figure}
\plotone{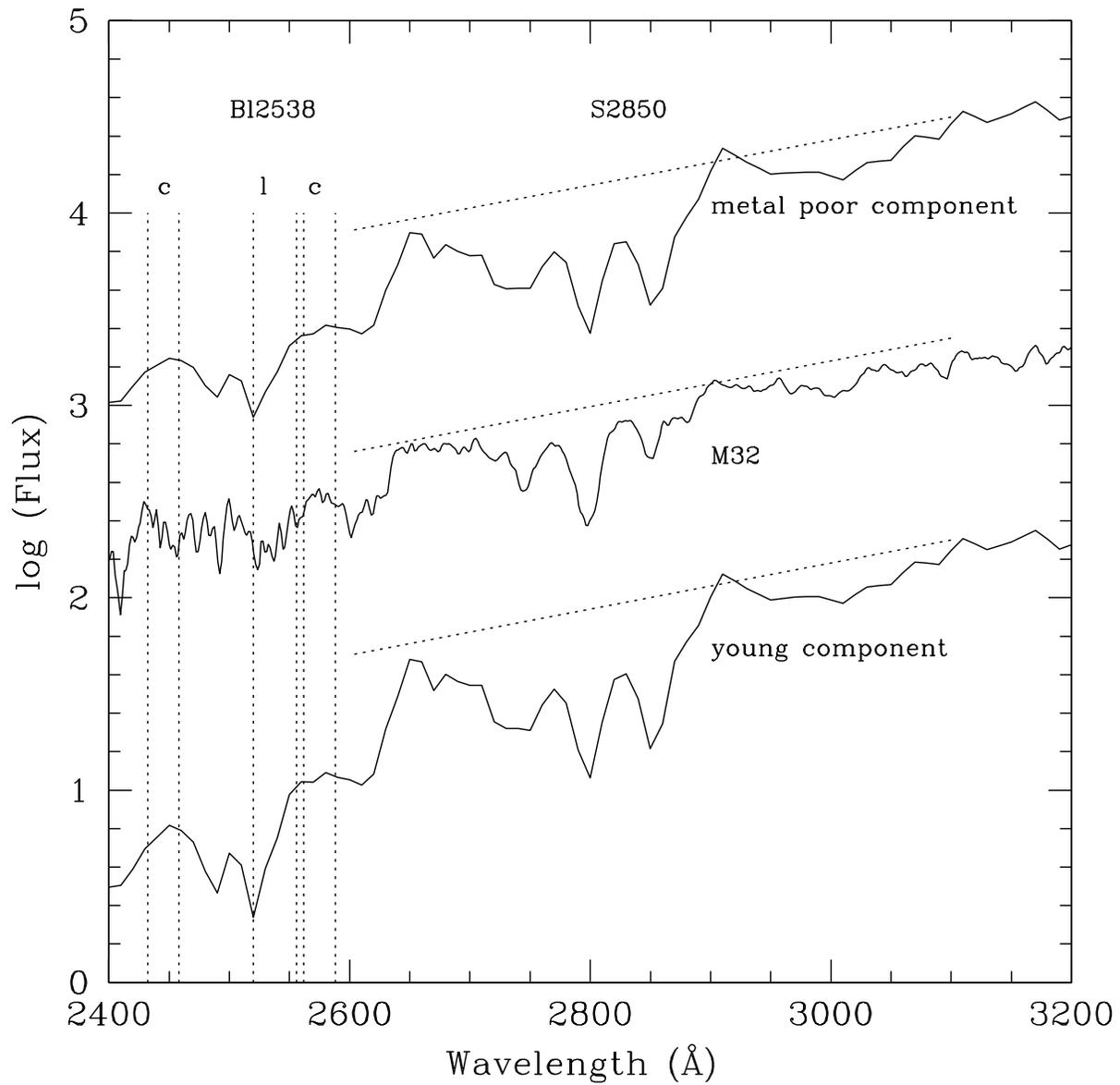}
\caption{
M 32 mid-UV spectra is compared to the 0.5 \% Z=0.0004/ 99.5\% Z=0.05
12 Gyr and the 5\% 3 Gyr / 95\% 12 Gyr Z=0.05 models. See Fig. 14a and Fig 15
for $Bl2538$ and $S2850$.}
\end{figure}

\end{document}